\documentclass[11pt,english]{article}

\usepackage{amsfonts,a4wide}
\usepackage{amsmath}
\usepackage{amssymb}
\usepackage{graphicx}
\usepackage{bm}
\usepackage{subfigure}
\usepackage{color}
\usepackage{xcolor}
\usepackage{tensor}
\usepackage{multirow}
\usepackage[normalem]{ulem}
\usepackage{hyperref}

\numberwithin{equation}{section}

\newcommand{\eqarray}[1]{\begin{eqnarray} #1 \end{eqnarray}}
\makeatother
\newlength{\x}
\newlength{\z}
\newlength{\abc}
\begin{document}

\title{\textbf{Spin oscillations of neutrinos scattered by the supermassive
  black hole in the galactic center}}

\author{Mridupawan Deka$^{a}$\thanks{mpdeka@theor.jinr.ru}\quad and\quad Maxim Dvornikov$^{a,b}$\thanks{maxim.dvornikov@gmail.com}
\\
\small{\ $^{a}$Bogoliubov Laboratory of Theoretical Physics,} \\
\small{Joint Institute for Nuclear Research,} \\
\small{141980 Dubna, Moscow region, Russia}
\\
\small{\ $^{b}$Pushkov Institute of Terrestrial Magnetism, Ionosphere} \\
\small{and Radiowave Propagation (IZMIRAN),} \\
\small{108840 Moscow, Troitsk, Russia}}

\date{}

\maketitle

\begin{abstract}
In this work, we study the propagation and spin oscillations of neutrinos in their scattering by a supermassive black hole (SMBH) surrounded by a realistic accretion disk. We review various descriptions of the fermion spin evolution in a curved spacetime under the influence of external fields. The overview of the test particle motion in the gravitational field of a rotating SMBH is also present. The external fields which a neutrino spin interacts with are the electroweak forces in plasma and the toroidal magnetic field in the accretion disk surrounding SMBH. Spin precession of neutrinos, which are supposed to be Dirac particles, is caused by the interaction of the neutrino magnetic moment with the magnetic field in the disk. We use a semi-analytical model of a thick accretion disk and review its characteristics. The cases of co-rotating and counter-rotating disks
with respect to BH are discussed.  We consider the incoming flux of neutrinos having an arbitrary angle with respect to the BH spin since the recent results of the Event Horizon Telescope indicate that the BH spin in the galactic center is not always perpendicular to the galactic plane. For our study, we consider a large number of incoming test neutrinos. We briefly discuss our results and their applications in the observations of astrophysical neutrinos.

\vspace*{5mm}

{\it Keywords}: neutrino spin oscillations; neutrino magnetic moment; accretion disk; astrophysical black hole; Kerr metric; magnetic fields.
\end{abstract}

\section{Introduction}
\label{sec:INTR}
Neutrinos are a unique tool to explore the laws of
  nature because their nonzero masses and mixing between
  flavors open a window to the physics beyond the standard
  model. The extremely weak opacity of matter for neutrinos
allows one to use such particles to probe various structures
which cannot be observed with the help of electromagnetic radiation.
This high penetrability is of great importance for the
neutrino astronomy. Sizable efforts are made in Refs.~\cite{All23,Abb24}
to detect neutrinos arriving on Earth from outside the solar
system. Moreover the attempts to observe multimessenger astronomy
effects involving neutrinos are undertaken,
e.g., in Ref.~\cite{Abb23}.

The observation of a neutrino in a detector is nontrivial
because of both rather weak strength of interaction with a
target and the additional feature which results from the fact
that neutrinos participate in electroweak interactions where
parity is violated. The neutrino interaction cross-section
depends on the neutrino helicity: left-handed neutrinos are
active whereas right-handed ones are sterile. Thus, if the spin
of a neutrino is flipped owing to the particle interaction with
external fields, we will observe the reduction of the initial
neutrino flux. If this process happens within one neutrino
generation, it is called neutrino spin oscillations. Originally,
neutrino spin oscillations were associated with non-trivial
neutrino electromagnetic properties, i.e. the possibility for a
neutrino to possess a nonzero magnetic
moment~\cite{Fujikawa:1980yx}. In this situation, the neutrino
spin-flip happens in an external magnetic field~\cite{Giu16}.

Other interactions, including the gravitational one, can also
contribute to the spin precession of a particle. The problem of
a spinning body motion in a gravitational field was studied,
first, in Ref.~\cite{Pap51}. The review of the extensive
literature on the classical and quantum approaches for the
description of spinning particles in gravitational fields and
non-inertial frames is present in Ref.~\cite{Ver23}.

Neutrino spin oscillations under the influence of a gravitational
field are studied in our works~\cite{Dvo06,Dvo13} based on the
quasiclassical approach. The formalism developed in
Refs.~\cite{Dvo06,Dvo13} is then applied in
Refs.~\cite{AlaNod15,Cha15,MasLam21,AlaFal22,PanMasLam22} to
study neutrino spin oscillations in various extensions of the
General Relativity (GR). The evolution of a high energy neutrino spin and helicity in various spacetimes using the quasiclassical approach was studied recently in Ref.~\cite{Saa25}.

In Refs.~\cite{Dvo23c,Dvo23d,Dvo23a,Dvo23b,Deka:2023ljj}, we
have applied the quasiclassical formalism for the studies of
spin effects in the neutrino gravitational scattering by a
black hole (BH) surrounded by a realistic magnetized accretion
disk. We have studied neutrino spin oscillations induced by poloidal
magnetic fields in Refs.~\cite{Dvo23c,Dvo23d}. Matter
effects have not been accounted in Refs.~\cite{Dvo23c,Dvo23d}
since the disk is considered
to be slim. We then take into account the electroweak interaction
of neutrinos with the plasma of the disk in Ref.~\cite{Dvo23a}.
However, in Ref.~\cite{Dvo23a}, we use a toy model of the disk
not substantiated by any calculations in relativistic MHD. More
realistic disk model, which is thick and contained inherent
magnetic fields, has been used in Ref.~\cite{Dvo23b} to study spin
oscillations in neutrino gravitational scattering. However, as
we shall discuss in the present work, the simulations in
Ref.~\cite{Dvo23b} had some numerical failures. Finally, we
have studied neutrino spin oscillations near BH with the help
of supercomputer simulations involving a large number of
scattered  test particles in Ref.~\cite{Deka:2023ljj}.

The motivations behind the studies of the spin effects in
the neutrino gravitational scattering are the following. First,
the neutrino in- and out-states are well defined
from the point of view of the quantum field theory in this kind
formulation of the problem. Second, this research has been
inspired by the recent observations of the silhouettes of the
event horizons of supermassive BHs (SMBH) in the centers of M87~\cite{Aki19}
and the Milky Way~\cite{Aki22}. This achievement of the
observational astronomy is a unique test of GR in the strong
field limit~\cite{DokNaz20}.

Some accretion disks around BHs are proposed in
Ref.~\cite{CheBel07} to emit high energy neutrinos. These
particles move in strong gravitational fields in the vicinity of
the BHs. The spin of such neutrinos can precess in the external
fields present in the accretion disk. It will lead to the
modification of the disk image seen in a neutrino telescope
compared to the optical one obtained, e.g., in
Ref.~\cite{BroCheMil97}. Indeed, some of the neutrinos can
become right-handed owing to spin oscillations. Thus, certain
regions of a disk will be invisible. We also mention that an extremely
high energy cosmic neutrino, which can be emitted in the vicinity
of a BH in the center of a blazar, has been observed recently~\cite{Aie25}.

Another possibility for the manifestation of the spin effects
in the neutrino gravitational scattering is when, e.g.,
supernova (SN) neutrinos are gravitationally lensed by the SMBH
in the center of our Galaxy. In this situation, we can observe
the reduction of the estimated neutrino flux if the Earth is at
certain position with respect to SN and the galactic center.
This kind of process is studied, e.g., in Ref.~\cite{MenMocQui06}.

In the present work, we continue our studies of spin oscillations
of neutrinos gravitationally scattered by a BH. Similar to
Refs.~\cite{Dvo23b,Deka:2023ljj}, we consider a rotating SMBH.
Following the  ``Polish doughnut'' model in
Ref.~\cite{Abramowicz_1978}, we introduce a thick magnetized
accretion disk surrounding the SMBH. In the present work, we
assume that the disk contains only the toroidal magnetic
field~\cite{Kom06} unlike in Ref.~\cite{Dvo23b,Deka:2023ljj}
where both toroidal and poloidal magnetic fields have been
considered together.

This work differs from our previous studies in the sense that
here we consider the incoming neutrino beams traversing
at a random angle with respect to the equatorial plane of the
SMBH instead of traveling parallel to it.
We consider a number of such incident angles in this study.
  This is motivated by the fact that the BH spin can be
  inclined towards the galactic plane in many cases. As an
  example, the spin of SMBH in our Galaxy is almost in the
  Galactic plane~\cite{Aki22V}. Since the relative position
  between a neutrino source and the Earth at the moment of the
  observation of neutrinos is not known, our study may shed
  light on the angle of inclination when neutrinos are
  gravitationally scattered by such a BH. Furthermore, it also
  allows us to investigate whether there exists an inverse
  symmetry in the observed fluxes of neutrinos for
  the opposite values of the cosine of the
  incident angles while all parameters are kept fixed. The
  absence of such symmetry shall reveal the inherent feature
  of the metric.

Moreover, we consider the cases where the accretion disk is
both co-rotating and counter-rotating with respect to the spin
of the BH. If we observe a clear difference between the
  observed fluxes of scattered neutrinos for both the disks
  with the same combination of BH spin and incident angle, this
  may help in inferring the direction of the disk rotation in the
measurement of astrophysical neutrinos.

Additionally in comparison to Ref.~\cite{Dvo23b}, we use a
few million incoming test neutrinos, with the help of High
Performance Parallel Computing, to significantly increase the
resolution of fluxes of the scattered neutrinos.

This work is organized as follows. We start in Sec.~\ref{sec:SPIN_EV} with the review of the fermion spin evolution in external fields, which include electromagnetic, electroweak, and gravitational interactions. These results are adapted for the description of the neutrino polarization in external backgrounds. In Sec.~\ref{sec:formalism},
we formulate the problem to be studied. In Sec.~\ref{sec:MOTION}, we review the description of a test particle trajectory in gravitational field of a rotating BH. In Sec.~\ref{sec:NUSPINEVOL}, the results of Secs.~\ref{sec:SPIN_EV} and~\ref{sec:MOTION} are applied for the spin evolution of neutrinos scattered by a rotating BH surrounded by a magnetized accretion disk. The properties of a particular accretion disk model are reviewed in
Sec.~\ref{sec:magnetized_accretion_disk}. The numerical parameters are
chosen in Sec.~\ref{sec:NUMERICAL}. We discuss our results
in Sec.~\ref{sec:RES} with a conclusion of our work given in
Sec.~\ref{sec:CONCL}.

\section{Evolution of a fermion spin in external fields}
\label{sec:SPIN_EV}

In this section, we review the evolution of a fermion spin in various external fields. We start with the description of the spin of a charged particle in an external electromagnetic field in the flat spacetime. Then, we generalize the spin evolution equation to account for the electroweak and gravitational interactions. In the case when a particle moves in a curved spacetime, we also consider the nonlinear evolution of a particle polarization, which takes place if the deviation from the geodesic motion is accounted for.

The polarization of an elementary particle is an essentially quantum characteristic. Indeed in the nonrelativistic quantum mechanics, the spin operator of a fermion, $\hat{\mathbf{s}} = \tfrac{\hbar}{2} \bm{\sigma}$, where $\bm{\sigma}=(\sigma_{1},\sigma_{2},\sigma_{3})$ being the Pauli matrices, contains the Planck constant $\hbar$. Nevertheless, the spin evolution enables the interpretation in terms of the classical physics.

If we consider a spinning particle with the magnetic moment $\mu$ in the external magnetic field $\mathbf{B}$, the interaction Hamiltonian has the form,
\begin{equation}
  H_\mathrm{int} = - \mu (\bm{\sigma} \cdot \mathbf{B}),
\end{equation}
The quantum Heisenberg equation for $\hat{\mathbf{s}}$ is $\tfrac{\mathrm{d}}{\mathrm{d}t}\hat{\mathbf{s}} = \tfrac{\mathrm{i}}{\hbar}[H_\mathrm{int},\hat{\mathbf{s}}]$. It eventually gives us,
\begin{equation}\label{eq:spinclass}
  \frac{\mathrm{d}}{\mathrm{d}t}\hat{\mathbf{s}} = 2 \mu (\hat{\mathbf{s}} \times \mathbf{B}),
\end{equation}
One can see in Eq.~\eqref{eq:spinclass} that the fermion spin precesses around the external magnetic field.

In Ref.~\cite{Bargmann:1959gz}, the result in Eq.~\eqref{eq:spinclass} is generalized for a particle, with the nonzero mass $m$, moving with an arbitrary velocity. For this purpose, we define the invariant spin $\bm{\zeta}$ in the particle rest frame. For a particle with an arbitrary momentum, the spin four vector is,
\begin{equation}\label{eq:spingen}
  S^\mu =
  (S^0,\mathbf{S}) =
  \left(
    \frac{\mathbf{p}\bm{\zeta}}{m},
    \bm{\zeta} + \frac{\mathbf{p}(\mathbf{p}\bm{\zeta})}{m(E+m)}
  \right),
\end{equation}
where $E = \sqrt{m^2+p^2}$ is the particle energy. One can see in Eq.~\eqref{eq:spingen} that $S^0 \to 0$ and $\mathbf{S} \to \bm{\zeta}$ in the particle rest frame.


If a charged particle moves in an electromagnetic field $F_{\mu\nu}$, the spin vector  $S^\mu$ in Eq.~\eqref{eq:spingen} obeys the Bargmann-Michel-Telegdi (BMT) equation which was also derived in Ref.~\cite{Bargmann:1959gz},
\begin{eqnarray}
  \frac{\mathrm{d}S^\mu}{\mathrm{d}\tau}
  &=&
  2\mu F^{\mu\nu} S_\nu -
  2\mu' U^\mu F^{\nu\lambda} U_\nu S_\lambda,
  \label{eq:BMT}
\end{eqnarray}
where $\tau$ is the proper
time. The quantities, $\mu$ and $\mu' = \mu - \mu_{\mathrm{B}}$, are the magnetic and anomalous
magnetic moments, respectively, while $\mu_{\mathrm{B}} = \frac{e}{2m}
= 5.8\times 10^{-9}\,\text{eV}\cdot\text{G}^{-1}$ is the Bohr magneton.
$U^\mu = \tfrac{\mathrm{d}x^\mu}{\mathrm{d}\tau} = p^\mu/m$ is the particle four velocity which satisfies the Lorentz
equation,
\begin{eqnarray}
  \label{eq:Lor}
  m \frac{\mathrm{d}U^\mu}{\mathrm{d}\tau} &=&
  e F^{\mu\nu} U_\nu.
\end{eqnarray}
Since a neutrino is an electrically neutral particle, hence $\mu = \mu'$ i.e. the
magnetic moment is purely anomalous. Moreover, 
$\frac{\mathrm{d}U^\mu}{\mathrm{d}\tau} = 0$ in Eq.~\eqref{eq:Lor},
i.e. a neutrino moves along a straight line in the flat space-time.

Equation~\eqref{eq:BMT} is then generalized to include the electroweak
interactions of neutrinos with the background matter in
Ref.~\cite{Dvornikov:2002rs} as follows,
\begin{eqnarray}
  \label{eq:BMTew}
  \frac{\mathrm{d}S^\mu}{\mathrm{d}\tau} &=&
  2\mu (F^{\mu\nu} S_\nu -
  U^\mu F^{\nu\lambda} U_\nu S_\lambda)
  +
  \sqrt{2}G_\mathrm{F}\varepsilon^{\mu\nu\lambda\rho} G_\nu U_\lambda S_\rho,
\end{eqnarray}
where $\varepsilon^{\mu\nu\lambda\rho}$ is the antisymmetric tensor in 
Minkowski space-time with the signature $\varepsilon^{0123} = 1$, and
$G_{\mathrm{F}}=1.17\times10^{-5}\,\text{GeV}^{-2}$ is the Fermi constant.
The quantity, $G^\mu$, is the contravariant effective potential for the
neutrino electroweak interactions with a background matter defined as, 
\begin{eqnarray}
  \label{eq:Gmu_1}
  G^\mu &=& \displaystyle\sum_f
  \left(
    q_f^{(1)} j_f^\mu + q_f^{(2)} \lambda_f^\mu
  \right).
\end{eqnarray}
Here, $j_f^\mu$ is the hydrodynamic current of background fermions of the
type $f$, $\lambda_f^\mu$ is their polarization. The constants,
$q_f^{(1,2)}$, are given in Ref.~\cite{Dvornikov:2002rs} for
various neutrino types.

If we now consider a spinning particle in curved spacetime,
the evolution of the particle's four momentum in an external
gravitational field is given in Refs.~\cite{Pap51,Wald:1972sz}
\begin{eqnarray}
  \label{eq:Papeq}
  \frac{\mathrm{D}p^\mu}{\mathrm{d}\tau} = &
  -\displaystyle\frac{1}{2} R^\mu_{\,\,\nu\rho\lambda} U^\nu S^{\rho\lambda},
\end{eqnarray}
where $R^\mu_{\,\,\nu\rho\lambda}$ is the Riemann tensor and $\mathrm{D}p^\mu = p^\mu_{\,;\nu}\mathrm{d}x^\nu$ is the covariant differential. The spin-tensor $S_{\mu\nu}$ is
related to the spin four vector by
\begin{eqnarray}
  \label{eq:SSrel}
  S^\mu = - \frac{1}{m^2} E^{\mu\nu\lambda\rho}p_\nu S_{\lambda\rho},
\end{eqnarray}
where $E^{\mu\nu\lambda\rho} = \frac{\varepsilon^{\mu\nu\lambda\rho}}{\sqrt{-g}}$ is
the antisymmetric tensor in the curved space-time,
$g = \text{det}(g_{\mu\nu})$ for the given metric $g_{\mu\nu}$, and $m$ is the particle mass
which is the integral of motion: $m^2 = p^2$.

In Sec.~\ref{sec:MOTION}, we briefly study the situation of a rotating BH, with the Kerr metric of spacetime around it. We use the Boyer-Lindquist coordinates in Sec.~\ref{sec:MOTION}. Although the results of the present section are general, we implicitly assume that the world coordinates $x^\mu$, used here, are  the Boyer-Lindquist ones.

Equations~\eqref{eq:BMT} and~\eqref{eq:BMTew} in flat space-time, as well as
  Eqs.~\eqref{eq:Papeq} and~\eqref{eq:SSrel} in curved space-time, are derived
  in frames of the quasiclassical approach originally. However, the fermion spin,
  which is a quantum object, requires the quantum treatment based on the Dirac
  equation. The derivation of Eqs.~\eqref{eq:Papeq} and~\eqref{eq:SSrel} based on
  the wave equation for a fermion accounting for the interaction with a gravitational
  field is carried out in Refs.~\cite{Aud81,Rud81,OanKum23}.

It is clear from Eq.~\eqref{eq:SSrel} that $\mathrm{D}p^\mu \neq 0$ if
$R^\mu_{\,\,\nu\rho\lambda} \neq 0$. Moreover, the four velocity $U^\mu$ and the particle four momentum
$p^\mu$ are not collinear in the general situation. Thus, unlike in the case of flat space-time,
the motion of a spinning neutrino deviates from the geodesics in an external gravitational
field. It should be also mentioned that, in deriving of Eq.~\eqref{eq:Papeq} in, e.g.
Refs.~\cite{Pap51,Wald:1972sz}, one assumes that the size of the spinning body is finite.

The spin tensor evolution equation, obtained in Refs.~\cite{Pap51,Wald:1972sz}, reads
\begin{equation}\label{eq:PapeqS}
  \frac{\mathrm{D}S^{\mu\nu}}{\mathrm{d}\tau} =
  p^\mu U^\nu - U^\mu p^\nu .
\end{equation} 
If one studies the evolution of the fermion polarization and neglects the right hand
side of Eq.~\eqref{eq:PapeqS}, the spin vector defined in Eq.~\eqref{eq:SSrel} is
parallel transported along the geodesics. Then, a fermionic spin does not affect the
particle motion. Equations~\eqref{eq:Papeq} and~\eqref{eq:PapeqS} were applied in Ref.~\cite{Ham24} for the description of the neutrino spin-flavor precession, i.e. when both the helicity and the flavor change, in a curved spacetime.

Despite Eqs.~\eqref{eq:Papeq}-\eqref{eq:PapeqS} are formally valid for particles with a nonzero mass, the limit of Eqs.~\eqref{eq:Papeq}-\eqref{eq:PapeqS} for ultrarelatiistic spinning fermions is considered in Ref.~\cite{PlyFenSte15}, where such particles orbiting the rotating and nonrotatring BHs are studied. That result is important for our studies. Shortly in Sec.~\ref{sec:MOTION}, we consider realistic astrophysical neutrinos which are ultrarelativistic. The study of the spinning fermions scattered  by BH, based on Eqs.~\eqref{eq:Papeq}-\eqref{eq:PapeqS}, is beyond the scope of the present work. We plan to discuss such a problem elsewhere.

Then, the formalism where the spin, e.g. of a neutrino, is parallel transported along the geodesics is proposed in
Ref.~\cite{Khriplovich:1997ni}. These results are later confirmed in the works of
Refs.~\cite{SorZil07,Obukhov:2017avp} which use the quantum approach based on the Dirac
equation analysis in curved space-time.

Finally, the spin evolution of neutrinos, assuming that a particle velocity is parallel transported along the geodesics, is given in
Ref.~\cite{Dvo13} by generalizing Eq.~\eqref{eq:BMTew} in a
curved space-time as,
\begin{eqnarray}
  \label{eq:BMTewgrav}
  \frac{\mathrm{D}S^\mu}{\mathrm{d}\tau} &=&
  2\mu (F^{\mu\nu} S_\nu -
  U^\mu F^{\nu\lambda} U_\nu S_\lambda)
  +
  \sqrt{2}G_\mathrm{F} E^{\mu\nu\lambda\rho} G_\nu U_\lambda S_\rho.
\end{eqnarray}
The definition of
$G^\mu$ remains the same as in Eq.~\eqref{eq:Gmu_1}, $j^\mu_f = n_f U^\mu_f$, where
$n_f$ is the invariant number density of background fermions, and $U^\mu_f$ is
the their four velocity in the curved space-time.

It should be noted that the right hand side of Eq.~\eqref{eq:BMTewgrav} is linear
in the spin vector $S^\mu$. However, if one solves more fundamental
  Eqs.~\eqref{eq:Papeq} and~\eqref{eq:PapeqS} to obtain the corrections to geodesic
  motion that is linear in spin, such corrections to the geodesic would introduce a
  correction to Eq.~\eqref{eq:BMTewgrav} which is quadratic in spin.

Instead of dealing with the Eq.~\eqref{eq:BMTewgrav} in curved space-time, it is
rather convenient to reformulate it for the invariant 3D spin vector. For this purpose, we
first make a transformation to a locally Minkowskian frame,
$x_a = e\indices{_a^\mu} x_\mu$, where $e\indices{_a^\mu}(a = 0\cdots 3)$ are the
vierbein vectors satisfying the relations,
\begin{eqnarray}\label{eq:vierbein_v_1}
  e\indices{_a^\mu}  e\indices{_b^\nu} g_{\mu\nu} &=& \eta_{ab}, \hspace{2mm}
  e\indices{_\mu^a}  e\indices{_\nu^b} \eta_{ab} \,=\, g_{\mu\nu}.
\end{eqnarray}
Here, $\eta_{ab} = \text{diag}(1,-1,-1,-1)$, is the Minkowski metric tensor.

If we now make a boost to the particle rest frame, we can write the
polarization of a neutrino in terms of an invariant three vector
$\bm{\zeta}$; cf. Eq.~\eqref{eq:spingen}. The evolution of the neutrino polarization vector obeys,
\begin{equation}\label{eq:nuspinrot}
  \frac{\mathrm{d}\bm{\bm{\zeta}}}{\mathrm{d}t}=2(\bm{\bm{\zeta}}\times\bm{\bm{\Omega}}),
\end{equation}
where,
\begin{eqnarray}
  {\bm{\Omega}} &=&
  {\bm{\Omega}}^{\mathrm{g}}
  + {\bm{\Omega}}^{\mathrm{em}}
  + {\bm{\Omega}}^{\mathrm{matt}}. 
\end{eqnarray}
The gravitational, electromagnetic and electroweak
interactions, ${\bm{\Omega}}^{\mathrm{g,em,matt}}$, respectively, are given in
Refs.~\cite{Dvo23a,Dvo23b} as,
\begin{eqnarray}\label{eq:Omegagemmatt}
  {\bm\Omega}^{\mathrm{g}}
  &=& \frac{1}{2 U^t}
  \left[{\bm b}_g + \frac{1}{1 + u^0} ({\bm e}_g \times {\bm u}) \right],\nonumber\\
  {\bm\Omega}^{\mathrm{em}}
  &=& \frac{\mu}{U^t}
  \left[u^0 {\bm b} - \frac{\bm u (\bm u \bm b)}{1 + u^0} + ({\bm e} \times {\bm u})\right],\nonumber\\
  {\bm\Omega}^{\mathrm{matt}}
  &=&  \frac{G_{\mathrm{F}}\bm u}{\sqrt{2} U^t}
  \left(g^0 - \frac{(\bm g\bm u)}{1 + u^0}\right).
\end{eqnarray}
Here $u^a = (u^0, \bm u) = e\indices{^a_\mu} U^\mu$ are velocity vectors with
respect to the locally Minkowskian frame,
$U^\mu = \tfrac{\mathrm{d}x^\mu}{\mathrm{d}\tau}$ is the particle velocity
in world coordinates,
$G_{ab} = (\bm e_g, \bm b_g) = \gamma_{abc} u^c$,
$\gamma_{abc} = \eta_{ad} e\indices{^d_{\mu; \nu}} e\indices{_b^\mu} e\indices{_c^\nu}$
are the Ricci rotation coefficients. The semicolon stands for a covariant derivative. Moreover,
$f_{ab} = e\indices{_a^\mu} e\indices{_b^\nu} F_{\mu\nu} = (\bm e, \bm b)$ is
the electromagnetic field tensor in the locally Minkowskian frame, and 
$g^a = (g^0, \bm g) = e\indices{^a_\mu} G^\mu$.

The vectors ${\bm\Omega}^{\mathrm{g}}$, ${\bm\Omega}^{\mathrm{em}}$, and ${\bm\Omega}^{\mathrm{matt}}$ in Eq.~\eqref{eq:Omegagemmatt} are valid for particles with a nonzero mass. It means that the four velocity $U^\mu$ is finite in Eq.~\eqref{eq:Omegagemmatt}. We have already mentioned that we shall consider ultrarelativistic neutrinos in the present work. However, the components of ${\bm\Omega}$ are shown in Ref.~\cite{Dvo23c} to allow the smooth limit $m/E \to 0$. Hence the spin evolution of an ultrarelativistic neutrino is well defined. For example, the explicit values of the components of $G_{ab}$, $f_{ab}$, and $g^a$ in Appendix~\ref{app:explicit_form} are derived in the limit $m/E \to 0$.

The electromagnetic contribution to the neutrino spin precession vector, ${\bm\Omega}^{\mathrm{em}}$, has both the terms, magnetic $\propto \bm{b}$ and the electric one $\propto\bm{e}$. In fact, the explicit value of ${\bm\Omega}^{\mathrm{em}}$ depends on the magnetic field in the neutrino rest frame since the invariant neutrino spin $\bm{\zeta}$ is given in such a frame. This rest frame is defined with respect to the locally Minkowskian frame $x^a$. If one takes an arbitrary electromagnetic field $(\bm{e},\bm{b})$ and makes the boost to the particle rest frame, one gets exactly the value of ${\bm\Omega}^{\mathrm{em}}$ in Eq.~\eqref{eq:Omegagemmatt}. Despite a massless particle particle does not have a rest frame, we can still carry out the above calculations, assuming that the neutrino mass is small but nonzero, and, then, consider the limit $m/E \to 0$.

One should mention that we have to take into account that $u^a$ is no longer a constant vector when we solve Eqs.~\eqref{eq:nuspinrot} and~\eqref{eq:Omegagemmatt} in an external gravitational field. A neutrino propagates along a straight line in a locally Minkowskian frame only in some particular cases, e.g., when a neutrino orbits BH; cf. Refs.~\cite{Dvo06,Dvo13,AlaNod15,Cha15}. In a general situation, we should take that $u^a(t)$ in Eq.~\eqref{eq:Omegagemmatt} is time dependent, which can be found by solving the geodesics equation which can written in the form~\cite{Dvo06},
\begin{equation}
  \frac{\mathrm{d}u^a}{\mathrm{d}\tau} =
  G^{ab} u_b,
\end{equation}
where $G_{ab}$ is defined below Eq.~\eqref{eq:Omegagemmatt}.

However, the situations when a neutrino is captured by BH and rotates around are not very interesting from the phenomenological point of view. Even if the spin of such a neutrino is flipped, this effect cannot be detected by a remote observer. We can study the situation when a beam of neutrinos is scattered by BH. In this case, the polarizations of  incoming and outgoing particles, which are in the flat spacetime asymptotically, have certain values. This problem is studied in Refs.~\cite{Dvo23c,Dvo23d,Dvo23a,Dvo23b,Deka:2023ljj}, where we have considered the neutrino scattering by a rotating BH surrounded by a realistic accretion disk.
The flux of incoming neutrinos has been taken to be parallel to the equatorial plane moving both above and below this plane.

If a particle orbits BH, its polarization is known to change owing to the gravity effects (see, e.g., Ref.~\cite{Sch60}). When we study the scattering of astrophysical neutrinos by BH, we should assume that these particles are ultrarelativistic. The question arises whether the polarization of an ultrarelativistic neutrino can change under the influence of a gravitational field if such a particle is scattered by BH. The authors of Refs.~\cite{Mer95,SinMobPap04} have argued that the polarization change is possible. However, the fact that a polarization of an ultrarelativistic fermion is conserved in a weak gravitational lensing is established in Ref.~\cite{LamPapPun05}. We have proved that the neutrino polarization remains unchanged for the arbitrary scattering by a nonrotaring BH~\cite{Dvo20} and by a rotating BH in case when a neutrino moves in the equatorial plane~\cite{Dvo21}. The fact of the spin oscillations absence is established numerically in Refs.~\cite{Dvo23a,Dvo23b,Deka:2023ljj}. However, the general proof of this statement is not known yet.

\section{Formulation of the problem of the neutrino gravitational scattering}
\label{sec:formalism}

\begin{figure}[htbp]
\centering
\subfigure[]
 {\label{fig:tourus_2}
   \includegraphics[width=0.47\hsize]{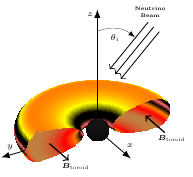}
 }
 \hspace{0mm}
 \subfigure[]
 {\label{fig:tourus_4}
   \includegraphics[width=0.47\hsize]{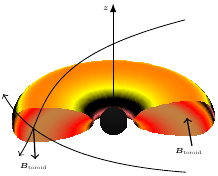}
 }
\protect 
\caption{Schematic diagrams showing neutrino trajectories.
  (a) A beam of neutrinos coming from $\infty$ at an angle
  $\theta_i$ with respect to the BH spin.
  (b) Neutrinos crossing the equatorial plane through
  the accretion disk.
  The toroidal magnetic field is perpendicular to their momenta.}
\end{figure}

We first consider that a beam of left-handed Dirac neutrinos is
emitted by a distant source. These neutrinos approach a spinning
SMBH with a polar angle $\theta_i$ with respect to the spin of the SMBH
such that their initial co-ordinates can be
written as $(r,\theta,\phi)_{\mathrm{s}} = (\infty,\theta_i,0)$;
cf. Fig.~\ref{fig:tourus_2}. Some of the neutrinos are captured
by the BH due to the gravitational pull. The rest are then
scattered gravitationally.

Note that we are not phenomenologically interested in
the situations when a neutrino is captured by the BH, and it
keeps rotating around it. Even if spin oscillations occur for
such a neutrino, it cannot be detected by a remote observer. We
can study the situation when a beam of neutrinos is scattered by
the BH, and one measures the polarization of incoming and
outgoing particles which are in the flat space-time
asymptotically.

Since we assume that a neutrino has a nonzero magnetic moment
$\mu$, the scattered beam of neutrinos experiences electromagnetic
interactions with the toroidal magnetic field of the accretion disk
(see Fig.~\ref{fig:tourus_4}), as it is described in Sec.~\ref{sec:SPIN_EV}. It also experiences electroweak interactions
with the background matter fields in the accretion disk. These interactions
cause neutrino spin precession resulting in spin oscillations. This implies
that some of the left-handed neutrinos become right-handed or sterile.

The gravitationally scattered neutrinos escape to infinity. Since some of
these neutrinos are likely sterile, not all of them can be observed at the
observer position,
$(r,\theta,\phi)_{\mathrm{obs}} = (\infty,\theta_\mathrm{obs},\phi_\mathrm{obs})$.
Our goal is to study the probability distributions of the left-handed neutrinos
remaining left handed as functions of $\theta_\mathrm{obs}$ and $\phi_\mathrm{obs}$.
In the following, we derive the trajectories and spin evolution of these
neutrinos between $(r,\theta,\phi)_{\mathrm{s}}$ and $(r,\theta,\phi)_{\mathrm{obs}}$. 

\section{Trajectories of ultra-relativistic neutrinos in Kerr Spacetime}
\label{sec:MOTION}

In this section, we describe the trajectories of ultra-relativistic
neutrinos in the presence of the gravitational field of a spinning BH. The
spacetime in this case is defined using the Kerr metric. In the Boyer-Lindquist coordinates,
$x^{\mu}=(t,r,\theta,\phi)$, the Kerr metric for a BH with mass $M$ and angular
momentum $J$ along the $z$-axis can be written as,
\begin{equation}
  \label{eq:Kerrmetr}
  \mathrm{d}s^{2} =
  g_{\mu\nu}\mathrm{d}x^{\mu}\mathrm{d}x^{\nu}=
  \left(
    1-\frac{rr_{g}}{\Sigma}
  \right)
  \mathrm{d}t^{2}+2\frac{rr_{g}a\sin^{2}\theta}{\Sigma}\mathrm{d}t\mathrm{d}\phi-\frac{\Sigma}{\Delta}\mathrm{d}r^{2}-
  \Sigma\mathrm{d}\theta^{2}-\frac{\Xi}{\Sigma}\sin^{2}\theta\mathrm{d}\phi^{2},
\end{equation}
where,
\begin{equation}
  \label{eq:dsxi}
  \Delta=r^{2}-rr_{g}+a^{2},
  \quad
  \Sigma=r^{2}+a^{2}\cos^{2}\theta,
  \quad
  \Xi=
  \left(
    r^{2}+a^{2}
  \right)
  \Sigma+rr_{g}a^{2}\sin^{2}\theta.
\end{equation}
Here, $r_g = 2M$ is the Schwarzschild radius, and $J = a M$ where $0 < a < M$.

The vierbein vectors, defined in Eq.~\eqref{eq:vierbein_v_1}, have the following explicit form for the Kerr metric in Eq.~\eqref{eq:Kerrmetr}:
\begin{eqnarray}
  e\indices{_0^\mu}
  &=& \left(
  \sqrt{\frac{\Xi}{\Sigma\Delta}}, 0, 0, \frac{a r r_g}{\sqrt{\Sigma\Delta\Xi}}
  \right), \hspace{2mm}
  e\indices{_1^\mu}
  \,=\, \left(0, \sqrt{\frac{\Delta}{\Sigma}}, 0, 0\right) , \nonumber\\
  e\indices{_2^\mu}
  &=& \left(0, 0, \frac{1}{\sqrt{\Sigma}}, 0 \right) , \hspace{2mm}
  e\indices{_3^\mu}
  \,=\, \left(0, 0, 0, \frac{1}{\sin\theta}\sqrt{\frac{\Sigma}{\Xi}}\right),
  \label{eq:vierbein_v_2}
\end{eqnarray}
where $\Sigma$, $\Delta$ and $\Xi$ are given in Eq.~\eqref{eq:dsxi}.

The geodesic motion of an ultra-relativistic test particle in Kerr metric has
three constants of motion:
\begin{itemize}
\item[]
  $E$: The total energy of the particle.
\item[]
  $L$: The projection of the angular momentum of the particle on the BH spin.
\item[]
  $Q$: The Carter constant.
\end{itemize}
In this study, $Q>0$ since, here, we are considering the scattering of particles
only~\cite{Cha83}. Moreover, we should take into account that $p^2 = m^2$
  for a neutrino with the nonzero mass $m$.
  While active neutrinos are well established to have small but nonzero
  masses in the sub-eV region, the typical energy of astrophysical neutrinos
  exceeds tens of MeV for SN neutrinos. Despite a massive particle moves along
  a time-like geodesics, we can assume with the high level of accuracy that the
  deviation of a trajectory from the null geodesics is negligible for a
  realistic astrophysical neutrino.

The trajectory of an ultra-relativistic particle in the presence of the
gravitational field of a spinning BH can be written in terms of two integral
equations as~\cite{GraLupStr18},
\begin{align}
  & \int\frac{\mathrm{d}r}{\pm\sqrt{R}}=\int\frac{\mathrm{d}\theta}{\pm\sqrt{\Theta}},
  \label{eq:trajth}
  \\
  & \phi = a\int\frac{\mathrm{d}r}{\pm\Delta\sqrt{R}}[(r^{2}+a^{2})E-aL]+
  \int\frac{\mathrm{d}\theta}{\pm\sqrt{\Theta}}
  \left[
    \frac{L}{\sin^{2}\theta}-aE
  \right],
  \label{eq:trajphi}
\end{align}
where $R$ and $\Theta$ potentials are defined as,
\begin{align}\label{eq:RTheta}
  R(r) = & [(r^{2}+a^{2})E-aL]^{2}-\Delta[Q+(L-aE)^{2}],
  \notag
  \\
  \Theta(\theta)= & Q + \cos^{2}\theta
  \left(
    a^{2} E^{2} - \frac{L^{2}}{\sin^{2}\theta} 
  \right). 
\end{align}
In Eqs.~\eqref{eq:trajth}-\eqref{eq:RTheta}, we take the limit $m/E \to 0$
since we consider the negligible deviation of a frajectory from a null
geodesics.

We choose $\pm{}$ signs for $\sqrt{R}$ and $\sqrt{\Theta}$ in
Eqs.~\eqref{eq:trajth} and~\eqref{eq:trajphi} to be the same as those of
$\mathrm{d}r$ and $\mathrm{d}\theta$ depending upon whether a neutrino is 
approaching from or moving away to infinity, e.g. if a neutrino is
approaching from infinity towards the BH, we choose negative sign for the
$r$-integral. Similarly, we choose positive sign if it is moving away to
infinity. When the neutrino is approaching the BH, the value of $\theta$
either decreases (minimum being $0$) or increases (maximum being $\pi$).
We choose negative sign for the $\theta$-integral in the first case and
positive sign for the second case. For computational convenience, we designate 
the neutrinos in the first case as the ``Upper'' neutrinos. The neutrinos
in the second case is designated as the ``Lower'' neutrinos. This
kind of separation of particles is equivalent to the choice of
the initial condition.

For the purpose of computation, we now define the following dimensionless
variables such that,
\begin{eqnarray}
  r = xr_{g},\, L = yr_{g}E,\, Q = wr_{g}^{2}E^{2},\, a = zr_{g},\,
  \mathfrak{\tilde t} = \cos\theta .
  \label{eq:dim_less_vars}
\end{eqnarray}
From now on all our expressions will be dimensionless.

Using Eq.~\eqref{eq:dim_less_vars}, we can rewrite Eqs.~\eqref{eq:trajth}
and~\eqref{eq:trajphi} as,
\begin{align}
  & z \int\frac{\mathrm{d}x}{\pm\sqrt{R(x)}}
  =
  \int\frac{\mathrm{d}\mathfrak{\tilde t}}{\pm\sqrt{\Theta(\mathfrak{\tilde t})}},
  \label{eq:trajth_1}
  \\
  & \phi
  = z\int\frac{(x-zy)\mathrm{d}x}{\pm\sqrt{R(x)}(x^{2}-x+z^{2})}
  + \frac{y}{z}\int\frac{\mathrm{d}\mathfrak{\tilde t}}{\pm\sqrt{\Theta(\mathfrak{\tilde t})}
    (1-\mathfrak{\tilde t}^{2})},
  \label{eq:trajphi_1}
\end{align}
where
\begin{align}
  \label{eq:RTh}
  R(x) =&
  \left(
    x^{2}+z^{2}-yz
  \right)^{2}-(x^{2}-x+z^{2})
  \left[
    w+
    \left(
    z-y
    \right)^{2}
    \right], \\
  \Theta(\mathfrak{\tilde t}) =&(\mathfrak{\tilde t}_{-}^{2}+\mathfrak{\tilde t}^{2})
  (\mathfrak{\tilde t}_{+}^{2}-\mathfrak{\tilde t}^{2}),\\
  \mathfrak{\tilde t}_{\pm}^{2} =&
  \frac{1}{2z^{2}}\left[\sqrt{(z^{2}-y^{2}-w)^{2}+4z^{2}w}\pm(z^{2}-y^{2}-w)\right],
\end{align}
with $\mathfrak{\tilde t}_{\pm}^2>0$. The value of the variable, $\mathfrak{\tilde t}$, as a
function of $x$ in Eq.~\eqref{eq:trajth_1} oscillates between $\pm \mathfrak{\tilde t}_{+}$,
$- \mathfrak{\tilde t}_{+} < \mathfrak{\tilde t} < + \mathfrak{\tilde t}_{+}$.

\begin{figure}[htbp]
\centering
\subfigure[]
 {\label{fig:bh_shadow_z0.01}
   \includegraphics[width=0.47\hsize]{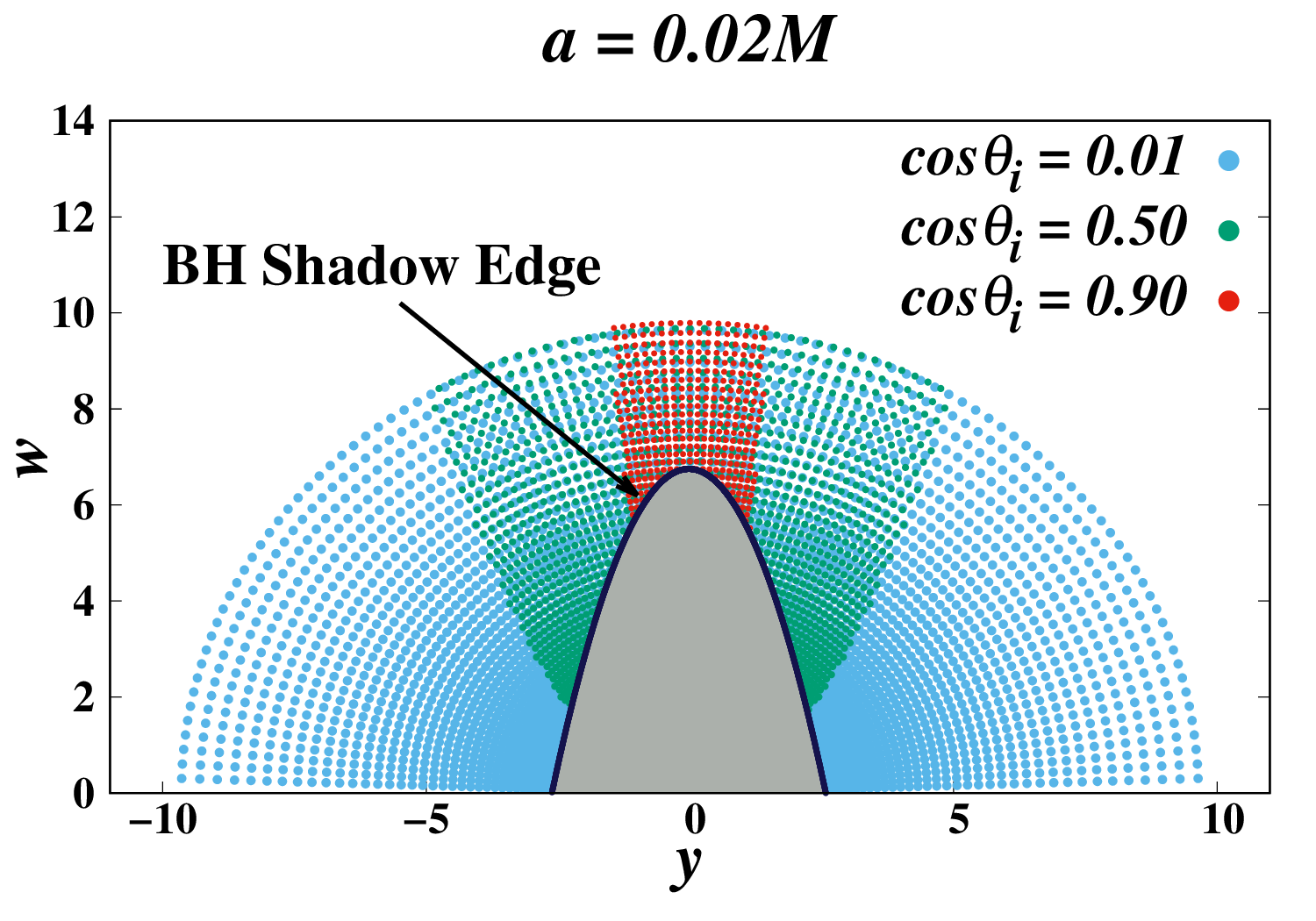}
 }
 \hspace{0mm}
 \subfigure[]
 {\label{fig:bh_shadow_z0.49}
   \includegraphics[width=0.47\hsize]{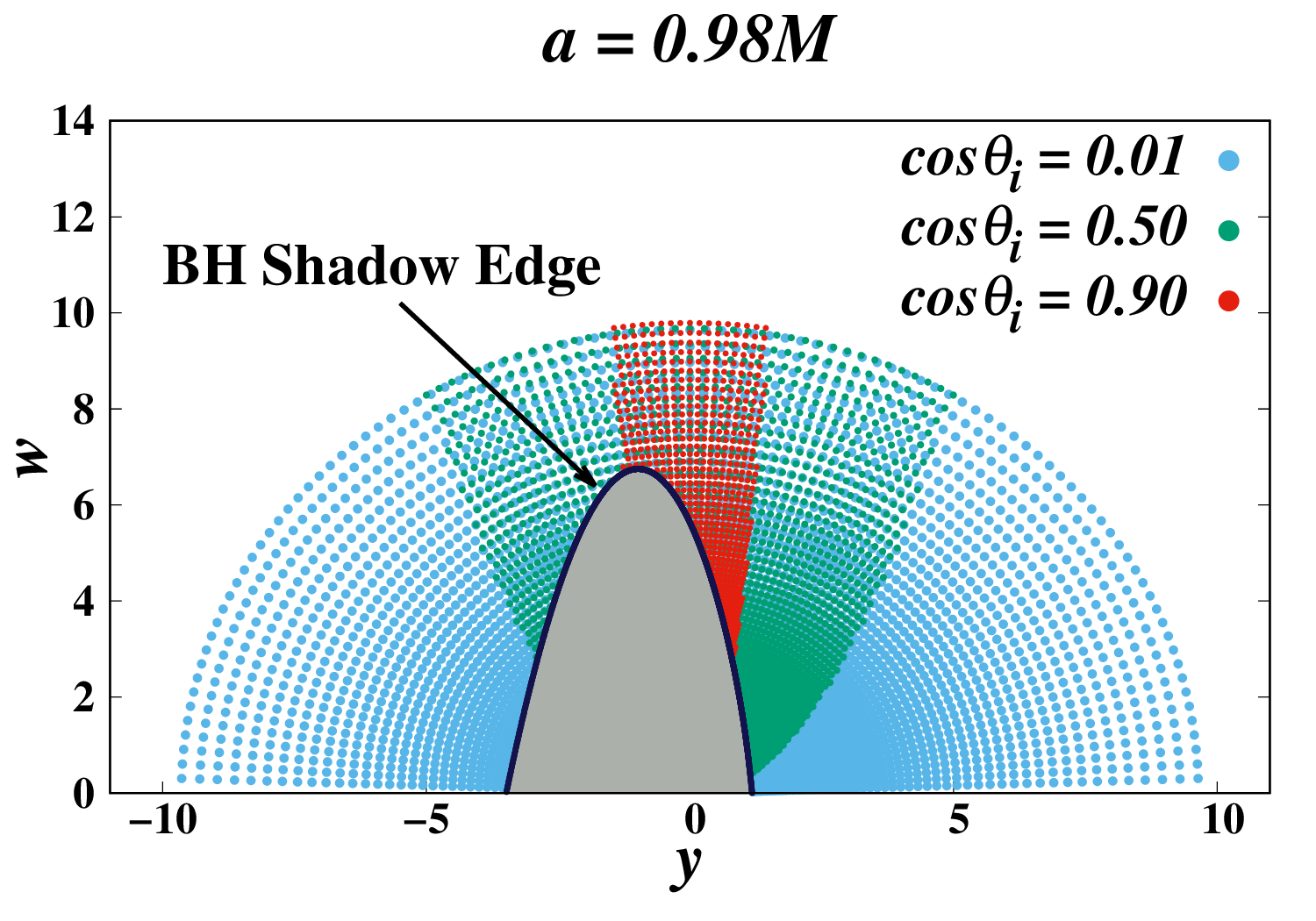}
 }
\protect 
\caption{Figures showing the dimensionless values of the integrals of motion,
  $y$ and $w$, together with the BH shadow. The points are spherically
  distributed in $y$ and $w$. They represent neutrino beams either coming out
  of or going into the page. We consider three different values of
  $\mathfrak{\tilde t}_i = \cos\theta_i$, namely
  $\mathfrak{\tilde t}_i = 0.01, 0.50$ and $0.90$ for each of the
  BH spins (a) $a = 0.02 M$ and (b) $a = 0.98M$. The beam or the point that
  falls in the shadow area is captured. For $\mathfrak{\tilde t}_i = 0.01$, only
  the neutrinos marked by light blue points are scattered. Similarly for
  $\mathfrak{\tilde t}_i = 0.50$ and
  $\mathfrak{\tilde t}_i = 0.90$, only the neutrinos marked by green and red points,
  respectively are scattered. In our computation, we consider only the scattered beams.
}
\end{figure}

Note that not all the trajectories defined by Eqs.~\eqref{eq:trajth}
and~\eqref{eq:trajphi} are going to survive since some of the particles
get captured by the BH and fall into it. The edge between the captured
and escaped particles satisfy the conditions, $R (\tilde r) = R' (\tilde r) = 0$.
These conditions lead to the parametric equations in terms of the following
dimensionless variables as,
\begin{eqnarray}
  \label{eq:y}
  y &=& - \frac{1}{z (2x - 1)}
  \left[x^2 (2x -3) + z^2 (2x + 1)\right],\\
  \label{eq:w}
  w &=& \frac{x^3}{z^2 (2x - 1)^2}
  \left[8z^2 - x (2x - 3)^2\right],\\
  \label{eq:rtilde}
  x_{\pm} &=& 1 + \cos\left[\frac{2}{3} \arccos(\pm 2z)\right],
  \hspace{2mm}
  x_{-} < x < x_{+}.
\end{eqnarray}
The parametric Eqs.~\eqref{eq:y}-\eqref{eq:rtilde}
define the BH shadow curve between the captured
and escaped particles. In Figs.~\ref{fig:bh_shadow_z0.01}
and~\ref{fig:bh_shadow_z0.49}, we show  BH shadow curves for the
two cases of BH spin, $a = 0.02 M$ and $a = 0.98M$, respectively.
The shadow areas seen by an observer can be reconstructed
using the results of, e.g., Ref.~\cite{GraLupStr18}.
The trajectories are either coming out of or going into the page.
  We use beams which are spherically distributed in $y$ and $w$.
  However, one may choose other distributions also. The trajectories,
  which fall inside the gray or the shadow 
  area, are the captured trajectories, and they discarded from the
  computation since we are interested in scattered neutrinos only.

In our previous works, e.g., in Ref.~\cite{Dvo23b}, we considered
the flux of incoming neutrinos which is parallel to the equatorial
plane. In the present work, we study the situation when the incoming
flux is still uniform. However, the angle between the axis of the
flux at $r\to\infty$ and the BH spin is arbitrary, $0<\theta_i<\pi$;
cf. Fig.~\ref{fig:tourus_2}. As a result, we shall have an additional
  condition on the allowed trajectories.

  Equations~\eqref{eq:y}-\eqref{eq:rtilde} guarantee that $\Theta(\mathfrak{\tilde t})>0$ in the integration in Eq.~\eqref{eq:trajth_1} if $\mathfrak{\tilde t}_i = \cos\theta_i = 0$, i.e. when the incoming flux is parallel to the equatorial plane. If $\mathfrak{\tilde t}_i \neq 0$, we should demand that
  $|\mathfrak{\tilde t}_{i}|<\mathfrak{\tilde t}_{+}$ to have $\Theta(\mathfrak{\tilde t})>0$ for
  any $\mathfrak{\tilde t}$ since the integration variable in Eq.~\eqref{eq:trajth_1} covers the segment $\mathfrak{\tilde t}_i < \mathfrak{\tilde t} < \mathfrak{\tilde t}_+$. Based on Eq.~\eqref{eq:RTh}, the condition $|\mathfrak{\tilde t}_i| < \mathfrak{\tilde t}_+$ can be rewritten as,
\begin{eqnarray}
  \label{eq:w_low_bound}
  w &>& - z^2 \mathfrak{\tilde t}_i^2 + y^2 \frac{\mathfrak{\tilde t}^2_i}{1 - \mathfrak{\tilde t}^2_i},
\end{eqnarray}
which should be considered along with Eqs.~\eqref{eq:y}-\eqref{eq:rtilde}. One can see from Eq.~\eqref{eq:w_low_bound} that, if $\mathfrak{\tilde t}_i = 0$, we have $-\infty< y < +\infty$ and $w>0$, as it should be when an incoming beam is parallel to the equatorial plane. If $\mathfrak{\tilde t}_i \neq 0$, the constraint in Eq.~\eqref{eq:w_low_bound} means that some of the values of $y$ and $w$ are not allowed in addition to the shadow area. 

\begin{figure}[htbp]
  \centering
  \includegraphics[width=0.47\hsize]{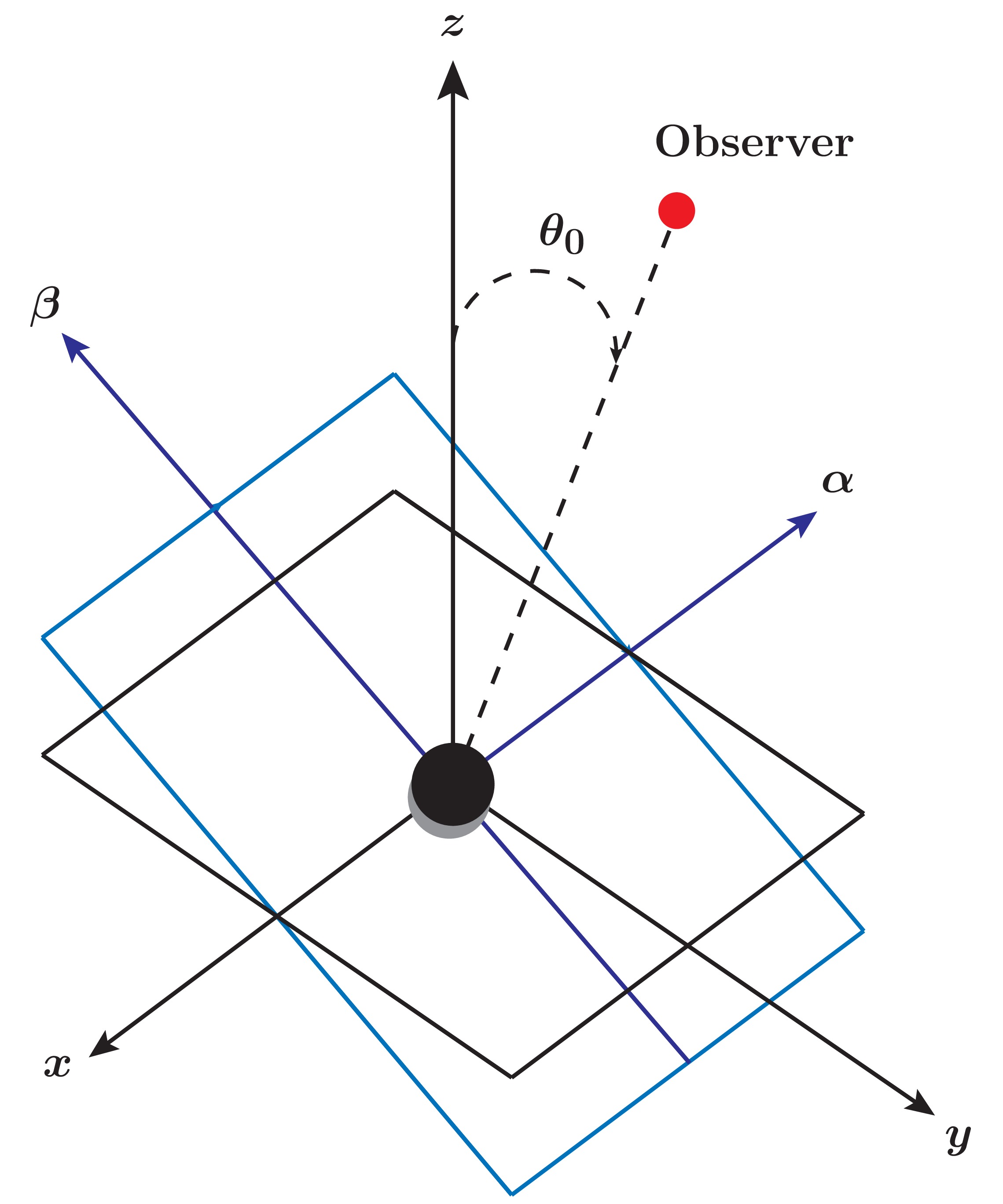}
  \protect 
  \caption{Figure showing the screen coordinates $\alpha$ and $\beta$. The observer
  is at an angle, $\theta_0$.}
  \label{fig:screen_coord}
\end{figure}

In Figs.~\ref{fig:bh_shadow_z0.01} and~\ref{fig:bh_shadow_z0.49}, we show the
accepted values of $y$ and $w$ for three different values of $\mathfrak{\tilde t}_i$,
i.e. $\mathfrak{\tilde t}_i = 0.01, 0.50$ and $0.90$.
When $\mathfrak{\tilde t}_i = 0.01$, we see that the accepted values of $y$ and $w$,
marked by light blue
  points, cover nearly the entire area except for the shadow area. However for
  $\mathfrak{\tilde t}_i = 0.50$, the accepted values of $y$ and $w$, marked by green points,
  are much less. Note that the green points overlap with the some of the blue points.
  For $\mathfrak{\tilde t}_i = 0.90$, the accepted values of $y$ and $w$, marked by red points,
  are even lower, and they overlap with some of both light blue and green points.

\begin{figure}[htbp]
\centering
\subfigure[]
 {\label{fig:screen_coord_shadow_z0.01}
   \includegraphics[width=1.\hsize]{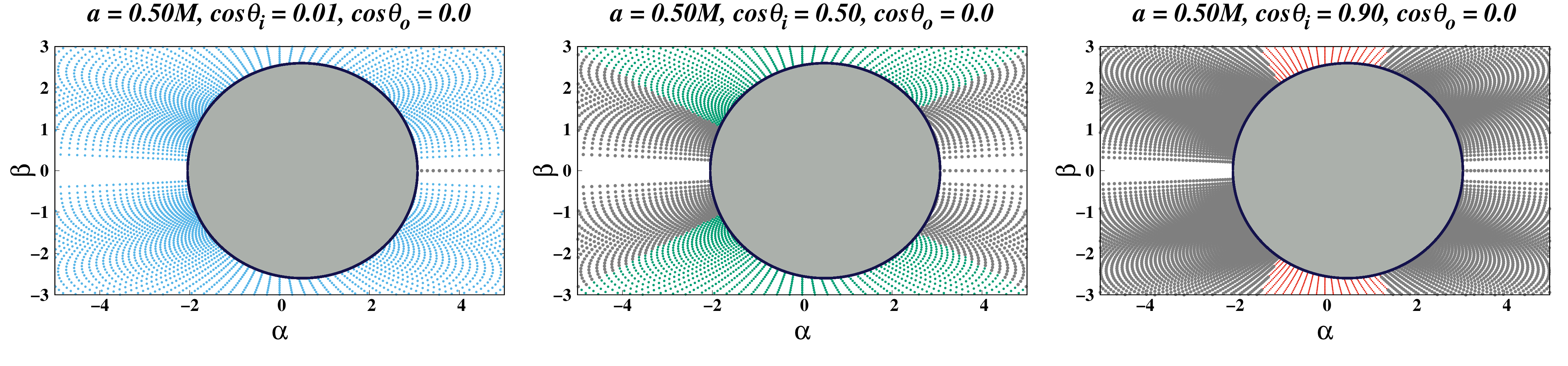}
 }
 \hspace{0mm}
 \subfigure[]
 {\label{fig:screen_coord_shadow_z0.49}
   \includegraphics[width=1\hsize]{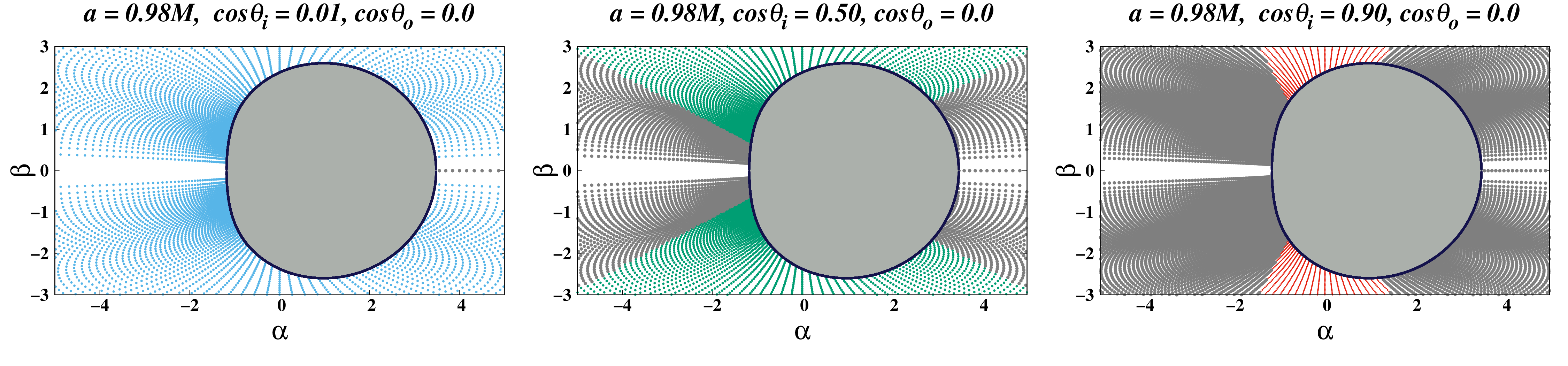}
 }
\protect 
\caption{Figures showing BH shadows in screen coordinates for two different
  BH spins, (a) $a = 0.50M$ and (b) $a = 0.98M$. For each BH spin, these figures
  represent different values of $\mathfrak{\tilde t}_i$, namely
  $\mathfrak{\tilde t}_i = 0.01, 0.50$ and $0.90$.
  The points represent neutrino beams either coming out of or going into
  the page.
  The gray colored beams (points) as well as the beams that are inside the BH
  shadow area are discarded from our computation.}
\end{figure}

Alternatively, one can also represent figures like Figs.~\ref{fig:bh_shadow_z0.01}
  and~\ref{fig:bh_shadow_z0.49} in terms of screen coordinates instead of $y$ and $w$.
  The screen coordinates $(\alpha, \beta)$ are defined in Ref.~\cite{GraLupStr18}, and
  shown in Fig.~\ref{fig:screen_coord}. The observer is at asymptotically flat
  space-time making an arbitrary polar angle, $\theta_0$. Then we can write the
  screen coordinates in terms of  $y$ and $w$ as,
\eqarray{
  \label{eq:scrn_coord}
  \alpha &=& - \frac{y}{\sin\theta_0}, \hspace{2mm}
  \beta \,=\, \pm \sqrt{w + 4 z^2 \cos^2\theta_0 - y^2 \cot^2\theta_0}, 
}
where, $\pm$ signs refer to the Northern and Southern hemispheres of the observer’s
  sky, respectively. In Figs.~\ref{fig:screen_coord_shadow_z0.01}
  and~\ref{fig:screen_coord_shadow_z0.49}, we
show the accepted and discarded beams as well as the BH shadow in terms of screen
coordinates for two BH spin, $a = 0.50M$ and $0.98M$. We take $\theta_0 = \pi/2$.
For each BH spin, these figures represent different values of $\mathfrak{\tilde t}_i$. The gray
colored beams (points) as well as the beams that are inside the BH shadow area are discarded.
Note that the white areas around $\beta = 0$ result from the relation
$\beta \sim \sqrt{w}$.

\subsection{Motion of the Upper Neutrinos}
\label{subsec:north_particles}

Let us first consider an upper neutrino approaching the BH from
infinity. As mentioned earlier, the value of the variable, $\mathfrak{\tilde t}$, as a
function of $x$ in Eq.~\eqref{eq:trajth_1} oscillates between $\pm \mathfrak{\tilde t}_{+}$.
We call the number of times when $t$ reaches either $- \mathfrak{\tilde t}_{+}$ or
$+ \mathfrak{\tilde t}_{+}$ by the number of the inversions of the trajectory, $N$. On
the other hand, we set the lower limit of $x$ to be $x > x_{\mathrm{tp}}$. $x_{\mathrm{tp}}$ is
the turn point which is the maximal real root of the equation $R(x) = 0$ with $R(x)$
being given in Eq.~\eqref{eq:RTh}. This is  the minimal distance to the BH center.

Generalizing the
results of Ref.~\cite{Deka:2023ljj}, we get the $N$ for incoming
neutrinos before they reach the turn point, $x_{\mathrm{tp}}$, as,
\begin{eqnarray}
  \label{eq:Ninv_in_north}
  N^{\mathrm{before}} &=&
  \left\lfloor \frac{I_x - F_i}{2 K} \right\rfloor +1,
\end{eqnarray}
where,
\begin{eqnarray}
  F_i &=& F \left(\arccos\frac{\mathfrak{\tilde t}_i}{\mathfrak{\tilde t}_+},
  \frac{\mathfrak{\tilde t}^2_+}{\mathfrak{\tilde t}^2_+ + \mathfrak{\tilde t}^2_-}\right),
\end{eqnarray}
is the Incomplete Elliptic integral of the first kind, $\lfloor\cdots\rfloor$
denotes the Floor function,
$K = K \left(\frac{\mathfrak{\tilde t}_{+}^{2}}{\mathfrak{\tilde t}_{-}^{2}+\mathfrak{\tilde t}_{+}^{2}}\right)$ is the
Complete Elliptic integral of first kind, and
\begin{equation}\label{eq:IxIt_in_north}
  I_x = z \sqrt{\mathfrak{\tilde t}_{+}^{2}+\mathfrak{\tilde t}_{-}^{2}} \int_x^\infty\frac{\mathrm{d}x'}{\sqrt{R(x')}},
\end{equation}
where $x_{\mathrm{tp}}< x < \infty$. For the definitions of Elliptic integrals and
functions, we follow those in Ref.~\cite{AbrSte64}.  Some of the useful
expressions are given in Appendix~\ref{app:elliptic_integrals}. The details of the calculations are provided in Appendix~\ref{sec:INTCALC}.

Using Eqs.~\eqref{eq:trajth_1} and~\eqref{eq:Ninv_in_north}, we can write
the relation for $\theta(x)$ for an incoming neutrino as,
\begin{eqnarray}
  \cos\theta_{\mathrm{in}} &=& \mathfrak{\tilde t}_+ \text{cn}
  \left(
  (-1)^{N^{\mathrm{before}}} \left\{
  F_i - I_x + 4K \left\lfloor \frac{N^{\mathrm{before}}}{2} \right\rfloor
  \right\}
  \bigg| \frac{\mathfrak{\tilde t}_{+}^{2}}{\mathfrak{\tilde t}_{-}^{2}+\mathfrak{\tilde t}_{+}^{2}}
  \right),
  \label{eq:thetabtp_north}
\end{eqnarray}
where $\text{cn}(n|m)$ is the Elliptic Jacobi function.

Similarly, using Eqs.~\eqref{eq:trajphi_1}, and~\eqref{eq:Ninv_in_north}, we can
write the corresponding relation for the azimuthal angle
$\phi_{\mathrm{in}}$ at the turn point as,
\begin{eqnarray}
  \phi_{\mathrm{in}} &=& z \displaystyle\int^\infty_{x_{\mathrm{tp}}}
  \frac{(x-zy) dx} {(x^2 -x + z^2) \sqrt{R(x)}}
  + \frac{y}{z\mathfrak{\tilde t}_-}
  \left\{2 N^{\mathrm{before}} \Pi - \Pi_i + (-1)^{N^{\mathrm{before}}} \Pi_{\mathrm{tp}}\right\},
  \label{eq:phibtp_north}
\end{eqnarray}
We do not need the $x$ dependence of $\phi(x)$ along the
whole trajectory. In Eq.~\eqref{eq:phibtp_north}, $\Pi$ is the Complete
Elliptic integral of third kind while $\Pi_i$ and $\Pi_{\mathrm{tp}}$
are the Incomplete Elliptic integrals of third kind. They are denoted as
follows,
\begin{equation}
  \Pi = \Pi\left({\mathfrak{\tilde t}}_{+}^{2}, - {\mathfrak{\tilde t}}_{+}^{2}/{\mathfrak{\tilde t}}_{-}^{2}\right),
  \Pi_i = \Pi\left( \arcsin({\mathfrak{\tilde t}}_i/{\mathfrak{\tilde t}}_+), {\mathfrak{\tilde t}}_{+}^{2}, - {\mathfrak{\tilde t}}_{+}^{2}/{\mathfrak{\tilde t}}_{-}^{2}\right),
  \Pi_{\mathrm{tp}} = \Pi\left(\arcsin({\mathfrak{\tilde t}}_{\mathrm{tp}}/{\mathfrak{\tilde t}}_+), {\mathfrak{\tilde t}}_{+}^{2}, - {\mathfrak{\tilde t}}_{+}^{2}/{\mathfrak{\tilde t}}_{-}^{2}\right).
\end{equation}
where
${\mathfrak{\tilde t}}_\mathrm{tp} = \cos\theta_{\mathrm{tp}}$ at the turn point
$x_{\mathrm{tp}}$.

When the neutrino escapes to infinity from the turn point, then the
number of inversions is given by,
\begin{eqnarray}
  \label{eq:Ninv_out_north}
\displaystyle N^{\mathrm{after}} &=&
  \begin{cases}
  \left\lfloor
    \displaystyle\frac{I_x + F_{\mathrm{tp}}}{2K} 
   \right\rfloor, & \text{if} \quad {\mathfrak{\tilde t}}'_\mathrm{tp} < 0,
   \\
   \\
  \left\lfloor
    \displaystyle\frac{I_x - F_{\mathrm{tp}}}{2K} 
   \right\rfloor + 1, & \text{if} \quad {\mathfrak{\tilde t}}'_\mathrm{tp} >0,
  \end{cases}
\end{eqnarray}
where $F_{\mathrm{tp}}$ is the Incomplete Elliptic integral of first kind denoted as, 
\begin{eqnarray}
  F_{\mathrm{tp}} &=& F
  \left(
  \arccos \frac{{\mathfrak{\tilde t}}_{\mathrm{tp}}}
          {{\mathfrak{\tilde t}}_{+}},\frac{{\mathfrak{\tilde t}}_{+}^{2}}
          {{\mathfrak{\tilde t}}_{-}^{2}+{\mathfrak{\tilde t}}_{+}^{2}}
  \right).
\end{eqnarray}
Note that, in Eq.~\eqref{eq:Ninv_out_north}, we distinguish the
cases whether the function ${\mathfrak{\tilde t}}(x)$ decreases or increases at the turn
point by considering the sign of
  ${\mathfrak{\tilde t}}' \equiv \tfrac{\mathrm{d}{\mathfrak{\tilde t}}}{\mathrm{d}x}$ at $x = x_\mathrm{tp}$. Note that, contrary to Eq.~\eqref{eq:IxIt_in_north}, we
  should take the limits of the $I_x$-integral for the particles
  moving from the turn point to infinity, e.g. in Eq.~\eqref{eq:Ninv_out_north}, as,
\begin{equation}\label{eq:IxIt_out_north}
  I_x =
  z \sqrt{{\mathfrak{\tilde t}}_{+}^{2}+{\mathfrak{\tilde t}}_{-}^{2}} \int_{{x}_\mathrm{tp}}^x\frac{\mathrm{d}x'}{\sqrt{R(x')}},
\end{equation}
where $x_{\mathrm{tp}}< x < \infty$.

Using Eqs.~\eqref{eq:trajth_1} and~\eqref{eq:Ninv_out_north}, we
can find $\theta(x)$ for the outgoing neutrinos as,
\begin{equation}
  \cos\theta_{\mathrm{out}} =
  {\mathfrak{\tilde t}}_+\times
  \begin{cases}
    \text{cn}
    \left(
      (-1)^{N^{\mathrm{after}}}
      \left(
        I_x + F_{\mathrm{tp}} - 4 K
        \left\lceil
          \displaystyle\frac{N^{\mathrm{after}}}{2}
        \right\rceil
      \right)
      \Big| \frac{{\mathfrak{\tilde t}}_{+}^{2}}{{\mathfrak{\tilde t}}_{-}^{2}+{\mathfrak{\tilde t}}_{+}^{2}}
    \right),
    & \text{if} \quad {\mathfrak{\tilde t}}'_\mathrm{tp} < 0,
    \\
    \text{cn}
    \left(
      (-1)^{N^{\mathrm{after}}}
      \left(
        F_{\mathrm{tp}} - I_x + 4 K
        \left\lfloor
          \displaystyle\frac{N^{\mathrm{after}}}{2}
        \right\rfloor
      \right)
      \Big| \frac{{\mathfrak{\tilde t}}_{+}^{2}}{{\mathfrak{\tilde t}}_{-}^{2}+{\mathfrak{\tilde t}}_{+}^{2}}
    \right),
  & \text{if} \quad {\mathfrak{\tilde t}}'_\mathrm{tp} >0.
  \end{cases}
  \label{eq:thetaatp_north}
\end{equation}
where, $\lceil\cdots\rceil$ denotes the Ceiling function.

Similarly using Eqs.~\eqref{eq:trajphi_1} and~\eqref{eq:Ninv_out_north},
we find $\phi_{\mathrm{out}}$ for an outgoing neutrino at the observer
position as,
\begin{equation}
  \phi_{\mathrm{out}} =
  \begin{cases}
    z \displaystyle\int^\infty_{x_{\mathrm{tp}}}
    \frac{(x-zy) dx} {(x^2 -x + z^2) \sqrt{R(x)}}
    + \frac{y}{z{\mathfrak{\tilde t}}_-}
    \left\{2 N^{\mathrm{after}} \Pi + \Pi_{\mathrm{tp}}
    - (-1)^{N^{\mathrm{after}}} \Pi_{{\mathfrak{\tilde t}}_{\mathrm{obs}}}\right\},
    & \text{if} \quad {\mathfrak{\tilde t}}'_\mathrm{tp} < 0,
    \\
    z \displaystyle\int^\infty_{x_{\mathrm{tp}}}
    \frac{(x-zy) dx} {(x^2 -x + z^2) \sqrt{R(x)}}
    + \frac{y}{z{\mathfrak{\tilde t}}_-}
    \left\{2 N^{\mathrm{after}} \Pi - \Pi_{\mathrm{tp}}
    + (-1)^{N^{\mathrm{after}}} \Pi_{{\mathfrak{\tilde t}}_{\mathrm{obs}}}\right\},
    & \text{if} \quad {\mathfrak{\tilde t}}'_\mathrm{tp} > 0.
    \label{eq:phiatp_north}
  \end{cases}
\end{equation}
Here,
\begin{eqnarray}
  {\mathfrak{\tilde t}}_{\mathrm{obs}} = \cos\theta_{\mathrm{obs}}, \hspace{2mm}
  \Pi_{{\mathfrak{\tilde t}}_{\mathrm{obs}}}
  = \Pi\left(\arcsin({\mathfrak{\tilde t}}_{\mathrm{obs}}/{\mathfrak{\tilde t}}_+), {\mathfrak{\tilde t}}_{+}^{2}, - {\mathfrak{\tilde t}}_{+}^{2}/{\mathfrak{\tilde t}}_{-}^{2}\right).
\end{eqnarray}

Note that $\phi_\mathrm{obs} = \phi_{\mathrm{in}} + \phi_{\mathrm{out}}$. Moreover,
since a neutrino can make multiple revolutions around the BH, this results in
$\phi_\mathrm{obs}$ being greater than $2\pi$. One should account for it in the final analysis. 

\subsection{Motion of the Lower Neutrinos}
\label{subsec:south_particles}

In a similar manner, we can also describe the trajectories of the Lower neutrinos.
We write the expressions for all the relevant quantities in the following.

Before the turn point,
\begin{eqnarray}
  \label{eq:Ninv_in_south}
  N^{\mathrm{before}} &=&
  \left\lfloor \frac{I_x + F_i}{2 K} \right\rfloor,
\end{eqnarray}
\begin{eqnarray}
  \cos\theta_{\mathrm{in}} &=& {\mathfrak{\tilde t}}_+ \text{cn}
  \left(
  (-1)^{N^{\mathrm{before}}} \left\{
  F_i + I_x - 4K \left\lceil \frac{N^{\mathrm{before}}}{2} \right\rceil
  \right\} \bigg| \frac{{\mathfrak{\tilde t}}_{+}^{2}}{{\mathfrak{\tilde t}}_{-}^{2}+{\mathfrak{\tilde t}}_{+}^{2}}
  \right),
  \label{eq:thetabtp_south}
\end{eqnarray}
\begin{eqnarray}
  \phi_{\mathrm{in}} &=& z \displaystyle\int^\infty_{x_{\mathrm{tp}}}
  \frac{(x-zy) dx} {(x^2 -x + z^2) \sqrt{R(x)}}
  + \frac{y}{z{\mathfrak{\tilde t}}_-}
  \left\{2 N^{\mathrm{before}} \Pi + \Pi_i - (-1)^{N^{\mathrm{before}}} \Pi_{\mathrm{tp}}\right\},
  \label{eq:phibtp_south}
\end{eqnarray}

After the turn point,
\begin{eqnarray}
  \label{eq:Ninv_out_south}
\displaystyle N^{\mathrm{after}} &=&
  \begin{cases}
  \left\lfloor
    \displaystyle\frac{I_x + F_{\mathrm{tp}}}{2K} 
   \right\rfloor, & \text{if} \quad {\mathfrak{\tilde t}}'_\mathrm{tp} < 0,
   \\
   \\
  \left\lfloor
    \displaystyle\frac{I_x - F_{\mathrm{tp}}}{2K} 
   \right\rfloor + 1, & \text{if} \quad {\mathfrak{\tilde t}}'_\mathrm{tp} >0,
  \end{cases}
\end{eqnarray}

\begin{equation}
  \cos\theta_{\mathrm{out}} =
  {\mathfrak{\tilde t}}_+\times
  \begin{cases}
    \text{cn}
    \left(
      (-1)^{N^{\mathrm{after}}}
      \left(
        I_x + F_{\mathrm{tp}} - 4 K
        \left\lceil
          \displaystyle\frac{N^{\mathrm{after}}}{2}
        \right\rceil
      \right)
      \Big| \frac{{\mathfrak{\tilde t}}_{+}^{2}}{{\mathfrak{\tilde t}}_{-}^{2}+{\mathfrak{\tilde t}}_{+}^{2}}
    \right),
    & \text{if} \quad {\mathfrak{\tilde t}}'_\mathrm{tp} < 0,
    \\
    \text{cn}
    \left(
      (-1)^{N^{\mathrm{after}}}
      \left(
        F_{\mathrm{tp}} - I_x + 4 K
        \left\lfloor
          \displaystyle\frac{N^{\mathrm{after}}}{2}
        \right\rfloor
      \right)
      \Big| \frac{{\mathfrak{\tilde t}}_{+}^{2}}{{\mathfrak{\tilde t}}_{-}^{2}+{\mathfrak{\tilde t}}_{+}^{2}}
    \right),
  & \text{if} \quad {\mathfrak{\tilde t}}'_\mathrm{tp} >0.
  \end{cases}
  \label{eq:thetaatp_south}
\end{equation}
\begin{equation}
  \phi_{\mathrm{out}} =
  \begin{cases}
    z \displaystyle\int^\infty_{x_{\mathrm{tp}}}
    \frac{(x-zy) dx} {(x^2 -x + z^2) \sqrt{R(x)}}
    + \frac{y}{z{\mathfrak{\tilde t}}_-}
    \left\{2 N^{\mathrm{after}} \Pi + \Pi_{\mathrm{tp}}
    - (-1)^{N^{\mathrm{after}}} \Pi_{{\mathfrak{\tilde t}}_{\mathrm{obs}}}\right\},
    & \text{if} \quad {\mathfrak{\tilde t}}'_\mathrm{tp} < 0,
    \\
    z \displaystyle\int^\infty_{x_{\mathrm{tp}}}
    \frac{(x-zy) dx} {(x^2 -x + z^2) \sqrt{R(x)}}
    + \frac{y}{z{\mathfrak{\tilde t}}_-}
    \left\{2 N^{\mathrm{after}} \Pi - \Pi_{\mathrm{tp}}
    + (-1)^{N^{\mathrm{after}}} \Pi_{{\mathfrak{\tilde t}}_{\mathrm{obs}}}\right\},
    & \text{if} \quad {\mathfrak{\tilde t}}'_\mathrm{tp} > 0.
    \label{eq:phiatp_south}
  \end{cases}
\end{equation}

Similar to the upper neutrinos,
$\phi_\mathrm{obs} = \phi_{\mathrm{in}} + \phi_{\mathrm{out}}$ for the
lower neutrinos too. Likewise since a neutrino can make multiple
revolutions around the BH, $\phi_\mathrm{obs}$ can be greater
than $2\pi$. One should adjust it in the final analysis.

The trajectories of the lower particles can be reconstructed from these for upper neutrinos if the incoming beam is parallel to the equatorial plane. However, we shall see soon that the spin evolution is different for these two groups of neutrinos especially in the presence of a magnetized accretion disk. Thus, one has to treat lower and upper particles independently. That is why we present the corresponding expressions here in the explicit form.

\begin{figure}[htbp]
\centering
\subfigure[]
 {\label{fig:r_vs_theta_z0.01}
   \includegraphics[width=0.47\hsize]{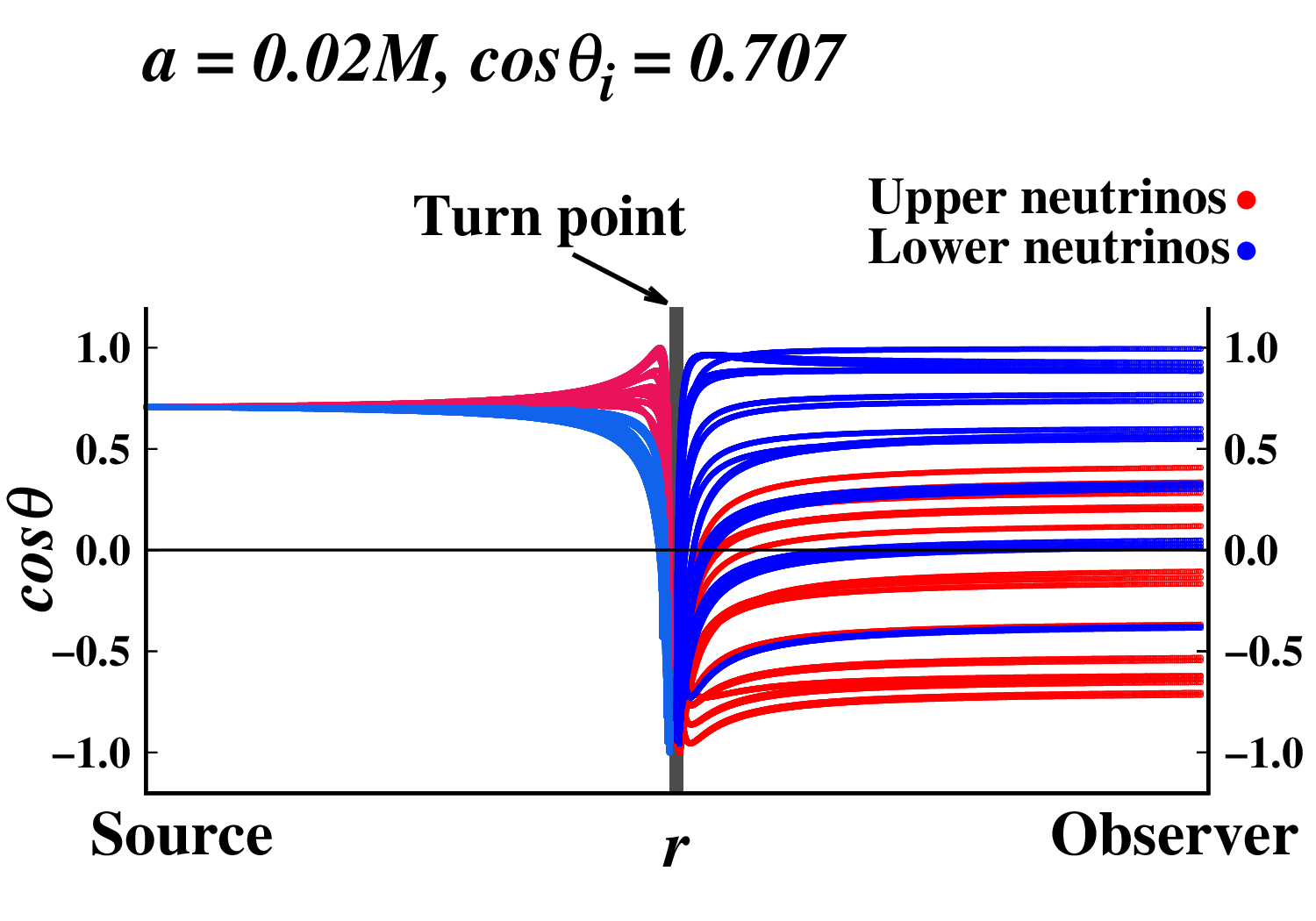}
 }
 \hspace{0mm}
 \subfigure[]
 {\label{fig:r_vs_theta_z0.49}
   \includegraphics[width=0.47\hsize]{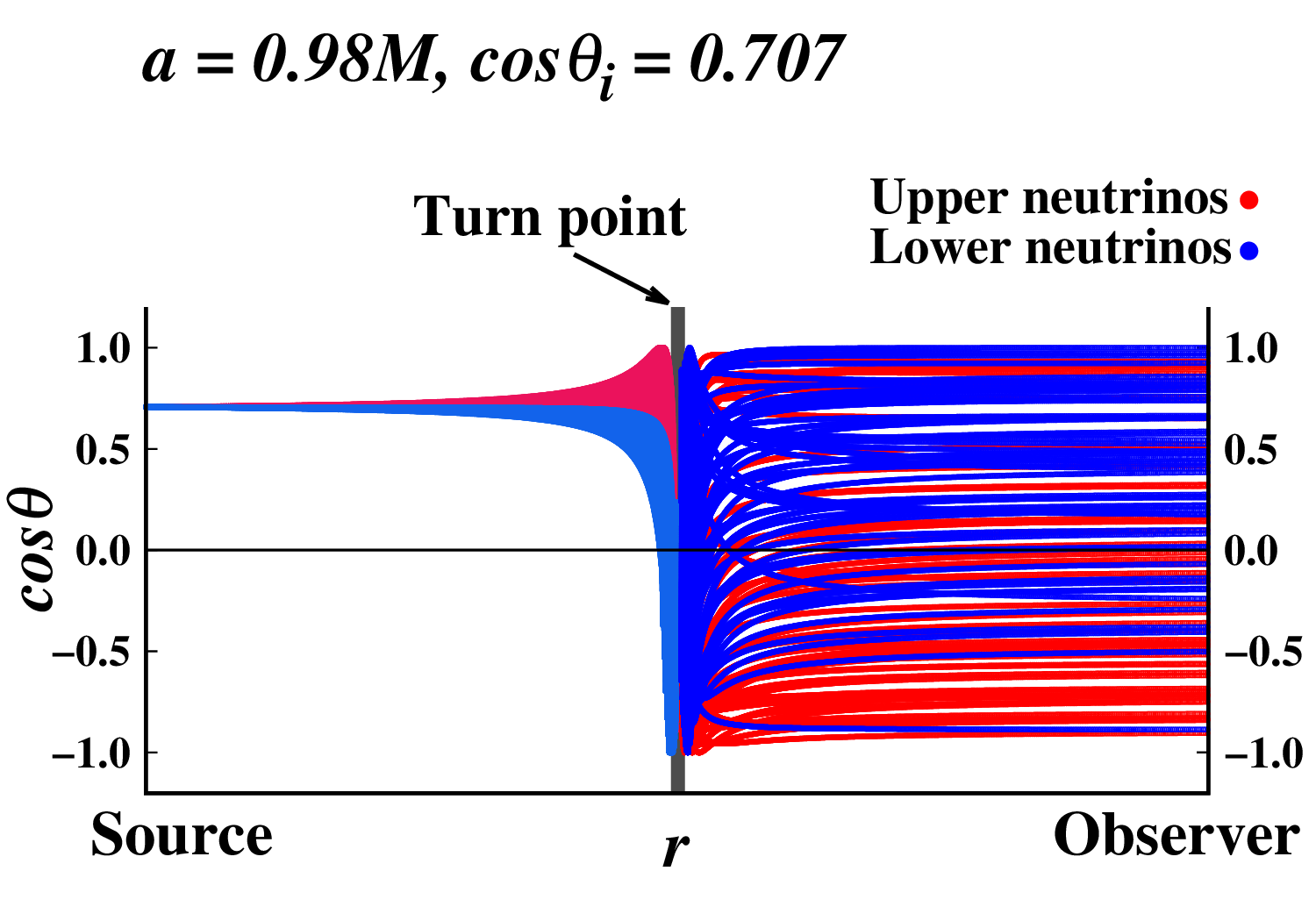}
 }
\protect 
\caption{Figures showing $r$ vs $\cos\theta$ for the trajectories
  of a few upper and lower neutrinos.
  (a) $a = 0.02 M$
  and
  (b) $a = 0.98M$. In both figures, the angle
  of incidence of the neutrino beam is $\theta_i = 45^{\circ}$, i.e.
  $\cos\theta_i = 0.707$. We can see that while approaching the BH,
  the neutrinos travel in straight line for most of the part. Only
  near the turn point, the beam separates into upper and lower
  neutrinos and start oscillating in $\cos\theta$. The $\cos\theta$
  behavour is dictated by the Eqs.~\eqref{eq:thetabtp_north},
  \eqref{eq:thetaatp_north}, \eqref{eq:thetabtp_south}
  and~\eqref{eq:thetaatp_south}. We see that the values of
  $\cos\theta$ experience changes in signs. This implies
  crossing of neutrinos from above the equatorial plane to below
  and vice versa.}
\end{figure}

In Figs.~\ref{fig:r_vs_theta_z0.01} and~\ref{fig:r_vs_theta_z0.49}, we
plot $\cos\theta$ versus
$r$  for the trajectories of a few upper and lower neutrinos for demonstration
purpose only with BH spins, $a = 0.02M$ and $a = 0.98M$, respectively. The angle
  of incidence of the neutrino beam is $\theta_i = 45^{\circ}$, i.e.
  $\cos\theta_i = 0.707$. The neutrinos are
moving in the pure gravitational field of the BH.
On their way to BH, for most of the part before the turn point,
  they travel in a straight line. Only near the turn point, the upper neutrinos
  move towards ${\mathfrak{\tilde t}}_+$. The maximum value of  ${\mathfrak{\tilde t}}_+$ is $1$, and some upper neutrinos
  reach this value. After reaching ${\mathfrak{\tilde t}}_+$, the upper neutrinos oscillate to some
  lower values, as low as $-1$. This change of signs in the $\cos\theta$ of the neutrinos
  implies the crossing of neutrinos from above the equatorial plane to below and vice versa.
  After crossing the turn point they again go through oscillations before escaping
  to the infinity. On the other hand, the lower neutrinos separate from the upper neutrinos
  and initially oscillate to $-{\mathfrak{\tilde t}}_-$ from ${\mathfrak{\tilde t}}_i = 0.707$. They also go through a few
  oscillations before escaping to infinity.

\section{Neutrino spin evolution in scattering by a rotating BH}\label{sec:NUSPINEVOL}

In Sec.~\ref{sec:SPIN_EV}, we consider the general evolution of a neutrino spin in a curved spacetime under the influence of electromagnetic and electroweak interactions. In this section, we apply these results for the situation of a rotating BH surrounded by a dense magnetized accretion disk.

The general expression for the vector $\bm{\Omega}$, driving the neutrino spin evolution, is provided in Eq.~\eqref{eq:Omegagemmatt}. In terms of the dimensionless variables in Eq.~\eqref{eq:dim_less_vars}, the quantities
${\bm{\Omega}}^{\mathrm{g,em,matt}}$ take the form as
\begin{eqnarray}
  \label{eq:omega_gravito_magnetic_dimensionless}
  \bm{\bm{\Omega}}_{x}^{\mathrm{g}}
  &=&\frac{1}{2}
  \left[
    \tilde{\bm{b}}_{g}+(\tilde{\bm{e}}_{g}\times\bm{v})
    \right],\\
  \label{eq:omega_em_dimensionless}
  \bm{\bm{\Omega}}_{x}^{\mathrm{em}}
  &=& V_{\mathrm{B}}
  \left[
    l^{0}\bm{b}-\bm{v}(\bm{lb})+(\bm{e}\times\bm{l})
    \right],\\
  \label{eq:omega_m_dimensionless}
  \bm{\bm{\Omega}}_{x}^{\mathrm{matt}}
  &=& V_{m}\bm{l}       
  \left[
    g^{0}-(\bm{gv})
    \right],
\end{eqnarray}
where the vectors, $(l^{0},\bm{l})=l^{a}=\frac{\mathrm{d}t}{\mathrm{d}r}\frac{u^{a}}{U^{t}}$,
$\bm{v}=\frac{\bm{u}}{1+u^{0}}$, $\tilde{\bm{e}}_{g}=\bm{e}_{g}\frac{r_{g}}{U^{t}}\tfrac{\mathrm{d}t}{\mathrm{d}r}$
and $\tilde{\bm{b}}_{g}=\bm{b}_{g}\frac{r_{g}}{U^{t}}\tfrac{\mathrm{d}t}{\mathrm{d}r}$,
are finite for an ultra-relativistic neutrino. Their explicit expressions
are given in Appendix~\ref{app:explicit_form}. The coefficients are
$V_{m} = \frac{G_{\mathrm{F}}}{\sqrt{2}m_p r^3_g}$ and
$V_B = \left(\frac{\mu}{r_g}\right)^2$, as well as $m_p$ is the mass of proton.

We should make a comment about the derivation of Eqs.~\eqref{eq:omega_gravito_magnetic_dimensionless}-\eqref{eq:omega_m_dimensionless}. The explicit form of $\bm{\Omega}$ 
is based on the particular vierbein vectors in Eq.~\eqref{eq:vierbein_v_2}. Studying the spin evolution of a particle moving in a weak gravitational field of a rotating body, it was shown in Ref.~\cite{ObuSilTer09} that the frequency of the spin precession depends on a tetrad used. There is a difference between the gravity contribution to the precession frequency found in Ref.~\cite{ObuSilTer09} and $\bm{\Omega}_\mathrm{g}$ in Refs.~\cite{Dvo06,Khriplovich:1997ni}.

The discrepancy found has the meaning of the relativistic corrections to the Lense-Thirring effect. It was mentioned in Ref.~\cite{ObuSilTer09} that probing this kind of corrections is beyond the sensitivity of the modern experiments.

Moreover, the neutrino invariant spin vector $\bm{\zeta}$ in Eq.~\eqref{eq:nuspinrot} is not an observable. Physics characteristics of a particle depend on its helicity $h = (\bm{\zeta}\cdot \bm{u}) / |\bm{u}|$. Thus, the study of the tetrad impact on the neutrino polarization should involve both the modification of Eqs.~\eqref{eq:omega_gravito_magnetic_dimensionless}-\eqref{eq:omega_m_dimensionless}, involving new vierbein vectors, and accounting for the mutual evolution of $\bm{\zeta}$ and $\bm{u}$. This issue is quite nontrivial for the case of the Kerr metric and is unlikely to be answered before explicit calculations are carried out. To the best of our knowledge there is no study of the impact of a tetrad on the spin related observables when a fermion interacts with a gravitational field. We are planning to examine whether the spin-flip probability depends on the tetrad in one of our future works.

The aim of the formalism developed in Refs.~\cite{Dvo06,Khriplovich:1997ni} consists in the obtaining the quasiclassical spin evolution equation with the right hand side linear in spin. This equation, including the chosen tetrad, was shown in Ref.~\cite{Khriplovich:1997ni} to give the correct results for nonrelativistic particles. If we study ultrarelativistic particles, such as astrophysical neutrinos, the formalism in Refs.~\cite{Dvo06,Khriplovich:1997ni} also predicts the correct helicity evolution in the neutrino gravitational scattering. It consists in the helicity conservation of a ultrarelativistic particle that was mentioned in Sec.~\ref{sec:SPIN_EV}. Depending on the complexity of a particular case including the situation of the Kerr metric, this statement was proven either analytically or numerically. In our opinion, these two features distinguish the tetrad choice in Refs.~\cite{Dvo06,Khriplovich:1997ni} among other possibilities.

One can associate the spin density matrix $\rho_\zeta = \tfrac{1}{2}(1+\bm{\sigma\zeta})$, where $\bm{\bm{\sigma}}=(\sigma_{1},\sigma_{2},\sigma_{3})$ are the Pauli matrices, with the invariant spin vector $\bm{\zeta}$. Since $\bm{\zeta}$ describes the polarization of a relativistic particle, the dynamics of $\rho_\zeta$ also corresponds to ultrarelativistic neutrinos. The quantum Liouville equation for $\rho_\zeta$ reads $\mathrm{i}\dot{\rho}_\zeta = [H,\rho_\zeta]_-$, where the Hamiltonian was shown in Ref.~\cite{LobPav99} to have the form,
\begin{equation}\label{eq:effHam0}
  \hat{H}= -\mathcal{U}_{2}(\bm{\bm{\sigma}}\cdot\bm{\bm{\Omega}})\mathcal{U}_{2}^{\dagger},
\end{equation}
where $\bm{\bm{\Omega}}$ is given in Eq.~\eqref{eq:Omegagemmatt}. In Eq.~\eqref{eq:effHam0}, we use the additional transformation in the spin space with the matrix $\mathcal{U}_{2} = \exp(\mathrm{i}\pi\sigma_{2}/4)$ since the quantization axis in the locally Minkowskian frame is $x^{a=1}$ rather than $x^{a=3}$ which is typically adopted in the majority of the spin evolution studies. This choice is because of the initial condition, which is discussed shortly.

If a neutrino is in a pure quantum spin state, we can associate the effective Schr\"odinger equation with the Hamiltonian in Eq.~\eqref{eq:effHam0}. It is convenient to rewrite this equation in the dimensionless variables as
\begin{equation}\label{eq:Schreq}
  \mathrm{i}\frac{\mathrm{d}\psi}{\mathrm{d}x}= \hat{H}_{x}\psi,
\end{equation}
where
\begin{equation}\label{eq:effHam}
  \hat{H}_{x}= -\mathcal{U}_{2}(\bm{\bm{\sigma}}\cdot\bm{\bm{\Omega}}_{x})\mathcal{U}_{2}^{\dagger},
  \quad
  \bm{\bm{\Omega}}_{x} =  r_{g}\bm{\bm{\Omega}}\frac{\mathrm{d}t}{\mathrm{d}r}.
\end{equation}
The explicit form of $\bm{\bm{\Omega}}_{x}$ is provided in Eqs.~\eqref{eq:omega_gravito_magnetic_dimensionless}-\eqref{eq:omega_m_dimensionless}. The effective wavefunction $\psi$ in Eq.~\eqref{eq:Schreq} describes the neutrino polarization. More detailed derivation of Eq.~\eqref{eq:Schreq} is provided in Ref.~\cite{LobPav99}.

Now we discuss the initial condition which Eq.~\eqref{eq:Schreq} should be supplied with. At $t \to -\infty$, incoming neutrinos have $\bm{u}_{-\infty} \propto (-1,0,0)$ in the locally Minkowskian frame since these particles move towards BH. Assuming that all incoming particles are left-handed, one gets that $\psi_{-\infty}^{\mathrm{T}}=(1,0)$. At $t \to +\infty$, outgoing neutrinos have $\bm{u}_{+\infty} \propto (+1,0,0)$ since these particles propagate from BH along the radial direction in the world coordinates. Hence $\psi_{+\infty}^{\mathrm{T}}=(\psi_{+\infty}^{(\mathrm{R})},\psi_{+\infty}^{(\mathrm{L})})$, where the components $\psi_{+\infty}^{(\mathrm{R,L})}$ are the complex numbers meaning the amplitudes to find neutrinos in the right- and left-polarized states. We recall that the physical state of a particle depends on the helicity $h = (\bm{u\zeta})/|\bm{u}|$. If the velocity $\bm{u}$ changes sign, the roles of the components in the effective wavefunction $\psi$ also interchange.

The Hamiltonian $\hat{H}_{x}$ is the function of $x$ only through the dependence
of $\theta(x)$ given in Eqs.~\eqref{eq:thetabtp_north}, \eqref{eq:thetaatp_north},
\eqref{eq:thetabtp_south} and~\eqref{eq:thetaatp_south}.
We solve the Eq.~\eqref{eq:Schreq} numerically at every $x$
separately for the upper and lower neutrinos. As we discussed earlier, the initial condition for Eq.~\eqref{eq:Schreq} is $\psi_{-\infty}^{\mathrm{T}}=(1,0)$, that corresponds to left polarized neutrinos at the infinity traveling at angle $\theta_i$ with respect to the spin of the BH.

We use discrete grid for the motion of the neutrinos. Our grid
is also irregular having denser points gradually towards the BH.
We first use $4$th order Adam-Bashforth predictor method for an
irregular grid to obtain the (approximate) solutions at
$x$. However as a neutrino approaches the BH, there is a
probability that we deal with a Stiff equation. In order to
avoid numerically unstable solutions, we then use Adam-Moulton
corrector method to iteratively improve the solutions obtained
from the Adam-Bashforth method with appropriate convergence
conditions.
The details of the  $4$th order Adam-Bashforth and Adam-Moulton
methods are provided in Appendix~\ref{app:ab_and_am_methods}.

There are two convergence condition that we use in the
Adam-Moulton method. The first one deals with the convergence
of each of the two components of $\psi(x)$, and the second one
checks whether the normalization condition is satisfied within
a particular iteration. In our case, these two conditions can
be written as,
\begin{eqnarray}
  \left|\frac{\psi^j(x)_k -  \psi^j(x)_{k-1}}
             {\psi^j(x)_{k-1}}\right| \leq 10^{-15},
             \hspace{2mm}
            \left|\psi(x)_k\right|^2 = 1 \pm O(10^{-15}), 
\end{eqnarray}
where, $k$ is the iteration number in the determination of
the $j$-th component of $\psi(x)$. When both the conditions
are satisfied, the solution is accepted.

After finding
$\psi_{+\infty}^{\mathrm{T}}=(\psi_{+\infty}^{(\mathrm{R})},\psi_{+\infty}^{(\mathrm{L})})$
numerically, we then compute
$P_{\mathrm{LL}} =|\psi_{+\infty}^{(\mathrm{L})}|^{2}$ at the observer
position. The angular co-ordinate $\theta_{\mathrm{obs}}$ can be
computed from Eqs.~\eqref{eq:thetaatp_north}
and~\eqref{eq:thetaatp_south}. Similarly, $\phi_{\mathrm{obs}}$ can
be computed from Eqs.~\eqref{eq:phibtp_north},
\eqref{eq:phiatp_north}, \eqref{eq:phibtp_south}
and~\eqref{eq:phiatp_south}.

\section{``Polish doughnut'' model of Magnetized Accretion Disk}
\label{sec:magnetized_accretion_disk}

To complete the neutrino spin evolution description we should specify the distributions of electromagnetic fields, as well as the density and the velocity of matter in an accretion disk. Namely, the vectors $g^a$, $\bm{e}$, and $\bm{b}$ in Eqs.~\eqref{eq:omega_em_dimensionless}
and~\eqref{eq:omega_m_dimensionless} are not defined yet. For this purpose we have to choose a particular accretion disk model.

There exists several models of accretion disk surrounding BH (For a
review, see e.g. Ref.~\cite{Abramowicz:2011xu}). For our study, when a neutrino
crosses the disk, its path should be long enough for the spin to rotate on a
sizable angle with respect to the neutrino velocity. For this reason,
we adopt a thick ``Polish doughnut'' disk proposed
in Ref.~\cite{Abramowicz_1978}. This model is later generalized in
Ref.~\cite{Kom06} to accommodate for the toroidal magnetic field. In this section, we briefly review the plasma and the magnetic field properties in frames of this model.

The plasma motion in an accretion disk is quite complex. We assume that the
hydrogen plasma is unpolarized, and we treat the
electroweak interaction of a neutrino with the fermions in the plasma in the
forward scattering approximation. We admit that the accretion disk can
rotate around BH with relativistic velocities.
We also assume that the plasma is electrically
neutral, i.e. the invariant number density of electrons and protons are equal,
$n_e = n_p$. The four potential, $G^\mu$, for the neutrino electroweak
interactions with a background matter can then be written as,
\begin{eqnarray}
  \label{eq:Gmu_2}
  G^\mu &=& \displaystyle\sum_{f = e,p}  q_f J_f^\mu,
\end{eqnarray}
where  $J_f^\mu = n_f U_f^\mu$ are the hydrodynamic currents and $U_f^\mu$ are
the four velocities of fermions in the disk. We assume that $U_e^\mu = U_p^\mu$,
i.e. there is no differential rotation between the components of the plasma.

The
only non-zero components in Eq.~\eqref{eq:Gmu_2} are $J_f^t = n_f U_f^t \neq 0$
and $J_f^\phi = n_f U_f^\phi \neq 0$ because of the axial symmetry of the disk.
Then the four potential in the locally Minkowskian frame,
$g^a = e\indices{^a_\mu} G^\mu$ can be written as,
\begin{eqnarray}
  g^a &=&
  \left(
    g^0,0,0, g^3
  \right).
  \label{eq:Gmu_3}
\end{eqnarray}
The explicit forms of $g^0$ and $g^3$ are provided in
Appendix~\ref{app:explicit_form}.

We can define the invariant extensions for the electric and magnetic fields in the plasma as,
\begin{eqnarray}
  E_\mu &=& F_{\mu\nu} U_f^\nu, \hspace{2mm}
  B_\mu \,=\, \tilde{F}_{\mu\nu} U_f^\nu, \hspace{2mm}
  \tilde{F}_{\mu\nu}= \frac{1}{2} E_{\mu\nu\alpha\beta}F^{\alpha\beta}.
\end{eqnarray}
Since the plasma is electrically neutral, $E_\mu = 0$. If we choose
$U_{f}^{\mu}=(U_{f}^{t},0,0,U_{f}^{\phi})$ and $B^{\mu}=(B^{t},0,0,B^{\phi})$,
then all the parameters of the disk depend on $r$ and $\theta$ due to the
axial symmetry of the metric in eq.~(\ref{eq:Kerrmetr}). Then
the only non-zero component of the  electromagnetic field tensor
$F^{r\theta} = - F^{\theta r}$. In that case, only the $b_3$ component of the toroidal
magnetic field in the local Minkowskian frame becomes non-zero. The explicit
form of $b_3$ is given in Appendix~\ref{app:explicit_form}.

We further assume that the specific angular momentum of a particle in the disk
$l = L/E$ is constant, $l=l_{0}$. Here $E = m U_{f}^{t}$ is the energy of the
particle, $L = - m U_{f}^{\phi}$ is its angular momentum, and $m$ is its mass.

Following Refs.~\cite{Abramowicz_1978,Kom06}, if we define the
generating function for the disk potential as,
\eqarray{
  \label{eq:genfunW}
  W(r,\theta) &=&
  \frac{1}{2}\ln
  \left|
  \frac{\mathcal{L}}{\mathcal{A}}
  \right|, 
}
then disk density $\rho$ and the magnetic pressure
$p_{m}^{(\mathrm{tor})}$ can be written as,
\eqarray{
  \label{eq:rhopm}
  \rho &=&
  \left[
    \frac{\kappa-1}{\kappa}
    \frac{W_{\mathrm{in}}-W}{K+K_{m}\mathcal{L}^{\kappa-1}}
    \right]^{\frac{1}{\kappa-1}},
  \hspace{2mm}
  p_{m}^{(\mathrm{tor})}\, =\,
  K_{m}\mathcal{L}^{\kappa-1}
  \left[
    \frac{\kappa-1}{\kappa}
    \frac{W_{\mathrm{in}}-W}{K+K_{m}\mathcal{L}^{\kappa-1}}
    \right]^{\frac{\kappa}{\kappa-1}},
}
where $K$, $K_{m}$, and $\kappa$ are the constants in the equations
of state. We choose $\kappa=4/3$ as in Ref.~\cite{Kom06}. The parameter
$W_{\mathrm{in}}$ in eq.~(\ref{eq:rhopm}) is the value of $W$ at the border of the disk. The maximal effective toroidal field
is given by $|\mathbf{B}|^{(\mathrm{tor})}_{\mathrm{max}}=\sqrt{2p_{m}^{(\mathrm{tor})}}$.

The components of $U_{f}^{\mu}$ and $B^{\mu}$ can then be written as,
\eqarray{
  \label{eq:UB}
  U_{f}^{t} &=&
  \sqrt{
  \left|
  \frac{\mathcal{A}}{\mathcal{L}}
  \right|
  }
  \frac{1}{1-l_{0}\Omega},
  \hspace{2mm}
  U_{f}^{\phi} \,=\, \Omega U_{f}^{t},
  \\
  B^{\phi} &=&\sqrt{\frac{2p_{m}^{(\mathrm{tor})}}{|\mathcal{A}|}},
  \hspace{2mm}
  B^{t} \,=\, l_{0}B^{\phi},
}
where,
\eqarray{
  \mathcal{L} &=& g_{tt}g_{\phi\phi}-g_{t\phi}^{2},\\
  \mathcal{A} &=& g_{\phi\phi}+2l_{0}g_{t\phi}+l_{0}^{2}g_{tt},\\
  \Omega &=& -\frac{g_{t\phi}+l_{0}g_{tt}}{g_{\phi\phi}+l_{0}g_{t\phi}}.
  \label{eq:Omegadisk}
}
Here, $\Omega$ is the angular velocity in the disk.

Equations~(\ref{eq:genfunW})-(\ref{eq:Omegadisk}) completely define all
the characteristics of the disk. For computation, we use the dimensionless
variables $\tilde{K}=r_{g}^{4(1-\kappa)}K$ and $\tilde{K}_{m}=r_{g}^{2(1-\kappa)}K_{m}$.
The methods to determine the values of $\tilde{K}$, $\tilde{K}_{m}$, $\rho$,
and $p_{m}^{(\mathrm{tor})}$ are discussed in Sec.~\ref{sec:NUMERICAL}

Note that the explicit value of the  specific angular momentum, $l_0$, depends
on whether the disk is co-rotating or counter-rotating~\cite{Bardeen:1972fi}.
The Keplerian angular momentum in terms of the dimensionless variable is given
by~\cite{Dvo23b,Kom06},
\eqarray{
  l(x)
  &=& \frac{\pm (x^{2} \mp z\sqrt{2x}+z^{2})}{\sqrt{2}x^{3/2}-\sqrt{2x}\pm z}.
  \label{eq:kepler_ang_mom}
}
where the upper sign is used for a co-rotating disk and the lower
sign for a counter-rotating one. Using Eq.~\eqref{eq:kepler_ang_mom}, one can
define two quantities $l_{\mathrm{mb}}$ and $l_{\mathrm{ms}}$ which are
the functions of the dimensionless radii of the marginally bound Keplerian
orbits, $x_{\mathrm{mb}}$, and, marginally stable Keplerian
orbits, $x_{\mathrm{ms}}$, respectively. The values of $x_{\mathrm{mb}}$
and $x_{\mathrm{ms}}$ also depend on whether the disk is co-rotating or
counter-rotating, and one shall compute them
accordingly~\cite{Bardeen:1972fi}. They are given as,
\eqarray{
  x_{\mathrm{mb}} &=& 1 \mp z + \sqrt{1 \mp 2  z},\\
  x_{\mathrm{ms}} &=& \frac{1}{2}[3 + Z_2 \mp \sqrt{ ( 3 - Z_1 ) ( 3 + Z_1 + 2 Z_2 ) } ],
}
where,
\eqarray{
  Z_1 &=& 1 + (1 - 4 z^2)^{1/3} [(1 + 2 z)^{1/3} + (1 - 2 z)^{1/3}],\\
  Z_2 &=& \sqrt{12 z^2 + Z_1^2}.
}
We then take the value of $l_0$ as,
\eqarray{
  l_0 &= & 0.6(l_{\mathrm{mb}}+l_{\mathrm{ms}}).
  \label{eq:l_0}
}
In our current work, we consider both  co-rotating or counter-rotating
disks.

\begin{table}[hbtp]
  \caption{The values of the parameters that are common to
    various combinations of BH spins and incident angles.}
  \label{tab:num_para_1}
  \begin{center}\setlength{\tabcolsep}{8pt}
    \begin{tabular}{c|c|c|c|c|c|c|c}
      \hline\hline
      {\bf BH mass}
      & {\boldmath $\mu$}    
      & {\boldmath $W_{\mathrm{in}}$}
      & {\boldmath $n_e^{\mathrm{max}}$}
      & {\boldmath $|\mathbf{B}|^{(\mathrm{tor})}_{\mathrm{max}}$}
      & {\boldmath $r_g$}
      & {\boldmath $V_{\mathrm{B}}$}
      & {\boldmath $V_m$}\\
      \hline\hline
      $10^8 M_\odot$
      & $10^{-13}\,\mu_\mathrm{B}$
      & $10^{-5}$
      & $10^{18}\,\text{cm}^{-3}$
      & $ 320\,\text{G}$
      & $3\times 10^{13}$ cm
      & $4\times 10^{-76}$
      & $2.5\times 10^{-87}$\\
      \hline\hline
    \end{tabular}
  \end{center}
\end{table}

\begin{table}[hbtp]
  \caption{The values of $\tilde{K}$ and $\tilde{K}_{m}$ for
    different BH spins for both the co-rotating and
    counter-rotating disks.}    
  \label{tab:num_para_2}
  \begin{center}\setlength{\tabcolsep}{8pt}
    \begin{tabular}{c|c|c|c}
      \hline\hline
      {\bf Disk type}
      & {\bf BH Spin}
      & {\boldmath $\tilde{K}$}
      & {\boldmath $\tilde{K}_{m}$}\\
      \hline\hline
      \multirow{3}{*}{\bf Co-rotating}
      & {$a = 0.02M$}
      & $2.55\times 10^{-31}$
      & $3.59\times 10^{-41}$\\
      \cline{2-4}
      & {$ a = 0.50M$}
      & $2.87\times 10^{-31}$
      & $3.01\times 10^{-41}$\\
      \cline{2-4}
      & {$ a = 0.98M$}
      & $9.35\times 10^{-31}$
      & $9.86\times 10^{-41}$\\
      \hline
      \multirow{3}{*}{\bf Counter-rotating}
      & {$a = 0.02M$}
      & $2.44\times 10^{-31}$
      & $3.62\times 10^{-41}$\\
      \cline{2-4}
      & {$ a = 0.50 M$}
      & $1.96\times 10^{-31}$
      & $3.12\times 10^{-41}$\\
      \cline{2-4}
      & {$ a = 0.98M$}
      & $1.66\times 10^{-31}$
      & $2.71\times 10^{-41}$\\
      \hline\hline
    \end{tabular}
  \end{center}
\end{table}

{
  \begin{table} 
    \caption{Number of gravitationally scattered
      neutrinos (in million) used for
      various combinations of BH spins and angles of incidence.
      The numbers for both the co-rotating and
      counter-rotating disks are listed.}
    \label{tab:number_of_neutrino}
    \begin{center}\setlength{\tabcolsep}{5pt}
      \begin{tabular}{c|c|c|c|c|c|c|c|c|c|}
      \hline\hline
      \multirow{2}{*}{\bf Disk type}
      & \multirow{2}{*}{\bf BH Spin}
      & \multicolumn{8}{|c|}
      {\bf{Number of Neutrinos (}{\boldmath{$\times 10^6$}} \bf{)}
        \bf{at} {\boldmath{$\cos\theta_i$ =}}}\\
      \cline{3-10}
      & & \boldmath{$0.01$}
      &  \boldmath{$- 0.01$}
      &  \boldmath{$0.50$}
      &  \boldmath{$- 0.50$}
      &  \boldmath{$0.707$}
      &  \boldmath{$- 0.707$}
      &  \boldmath{$0.90$}
      &  \boldmath{$- 0.90$}\\
      \hline\hline
      \multirow{3}{*}{\bf Co-rotating}
      & {$a = 0.02M$}
      & $2.68$
      & $2.68$
      & $3.20$
      & $3.20$
      & $2.32$
      & $2.32$
      & $2.68$
      & $2.68$\\
      \cline{2-10}
      & {$ a = 0.50M$}
      & $2.78$
      & $2.78$
      & $3.45$
      & $3.45$
      & $2.56$
      & $2.56$
      & $2.97$
      & $2.97$\\
      \cline{2-10}
      & {$ a = 0.98M$}
      & $2.91$
      & $2.91$
      & $4.04$
      & $4.00$
      & $3.24$
      & $3.25$
      & $3.99$
      & $4.00$\\
      \hline
      \multirow{3}{*}{\bf Counter-rotating}
      & {$a = 0.02M$}
      & $2.69$
      & $2.69$
      & $3.20$
      & $3.20$
      & $2.33$
      & $2.33$
      & $2.68$
      & $2.68$\\
      \cline{2-10}
      & {$ a = 0.50 M$}
      & $2.78$
      & $2.78$
      & $3.45$
      & $3.37$
      & $2.56$
      & $2.56$
      & $2.97$
      & $2.97$\\
      \cline{2-10}
      & {$ a = 0.98M$}
      & $3.32$
      & $3.32$
      & $4.83$
      & $4.83$
      & $3.75$
      & $3.75$
      & $4.21$
      & $4.21$\\
      \hline\hline     
      \end{tabular}
  \end{center}
\end{table}
}

\section{Numerical Methods and Parameters}
\label{sec:NUMERICAL}

In our study, we fix the mass of SMBH at $M = 10^8 M_\odot$. We consider three 
different spins of SMBH, e.g. $a\,=\,2\times 10^{-2} M, 0.50 M$ and
$0.98 M$. For each BH spin, we consider eight different incident angles,
 $\cos\theta_i = \pm 0.01, \pm 0.50, \pm 0.707$
  and $\pm 0.90$. We adopt a thick ``Polish doughnut" model for the accretion
disk~\cite{Abramowicz_1978}. The accretion disk consists of hydrogen plasma.
We consider both co-rotating and counter-rotating disks around the SMBH
with a relativistic velocity~\cite{Kom06}.

The magnetic moment of the Dirac neutrinos is taken to be
$\mu = 10^{-13}\,\mu_\mathrm{B}$ where $\mu_\mathrm{B}$ is the Bohr magneton. This
value is less than the
astrophysical upper bound of the magnetic moment~\cite{Via13}. We assume that the
neutrinos undergo electroweak interaction with the plasma of the accretion disk
in the forward scattering approximation~\cite{Dvornikov:2002rs}. We also assume
the neutrino spin oscillations within one neutrino generation, i.e. we suppose
that only the diagonal magnetic moment is present.

We set the value of the generating function, $W$, at the inner
border of the disk, $W_{\mathrm{in}} = 10^{-5}$. The values of the
parameters that are common to various BH spins and
incident angles are listed in Table~\ref{tab:num_para_1}.

\begin{figure}[htbp]
\centering
\subfigure[]
 {
   \includegraphics[width=0.49\hsize]{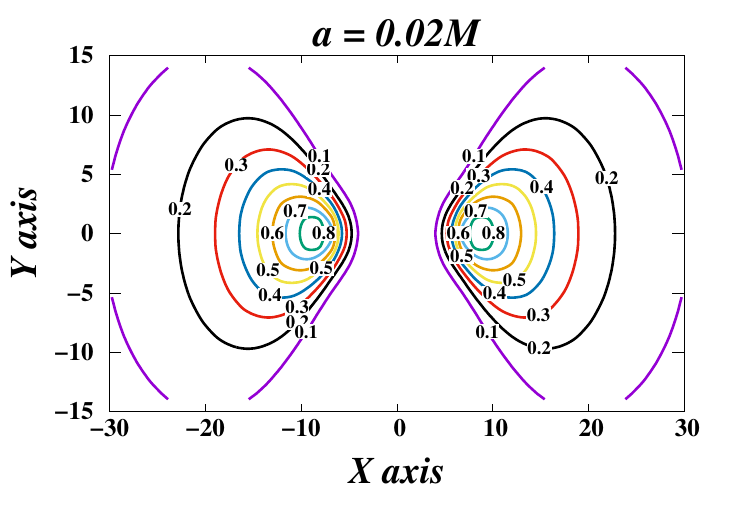}
 }
 \hspace{-5mm}
 \subfigure[]
 {
   \includegraphics[width=0.49\hsize]{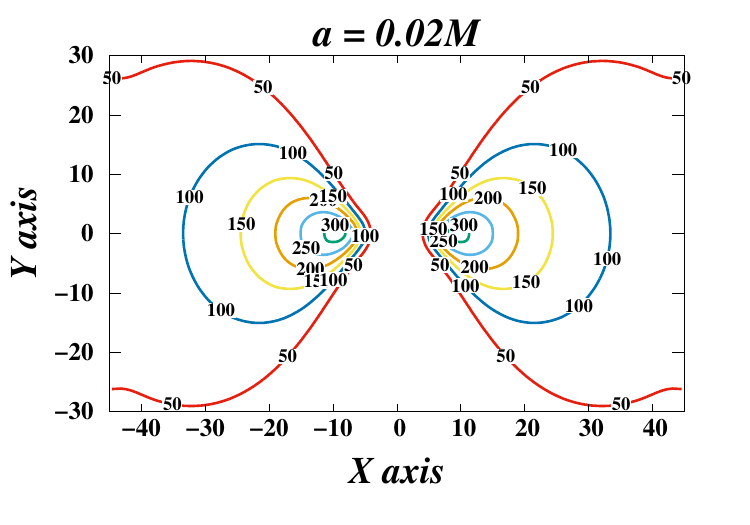}
 }
\subfigure[]
 {
   \includegraphics[width=0.49\hsize]{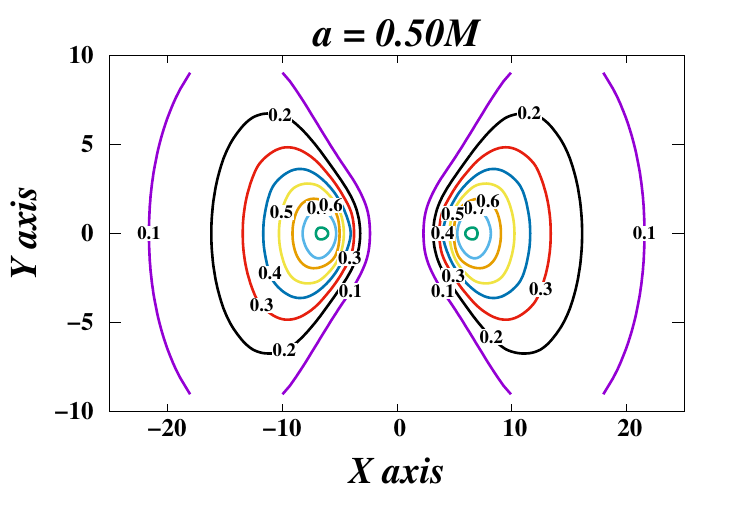}
 }
 \hspace{-5mm}
 \subfigure[]
 {
   \includegraphics[width=0.49\hsize]{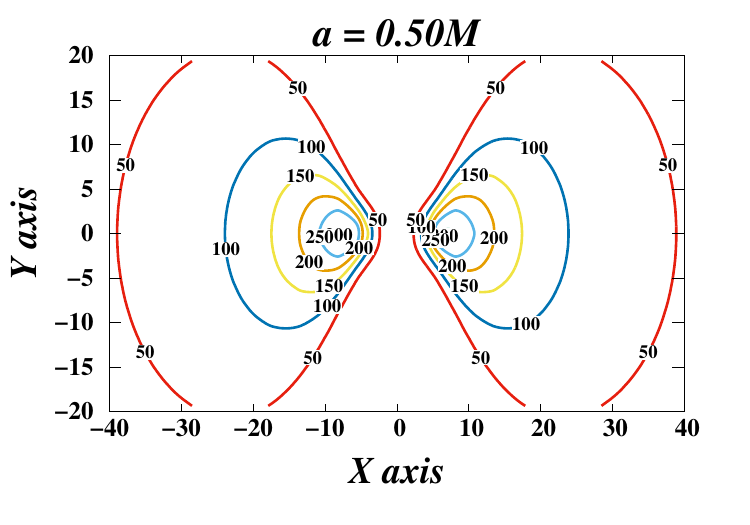}
 }
\subfigure[]
 {
   \includegraphics[width=0.49\hsize]{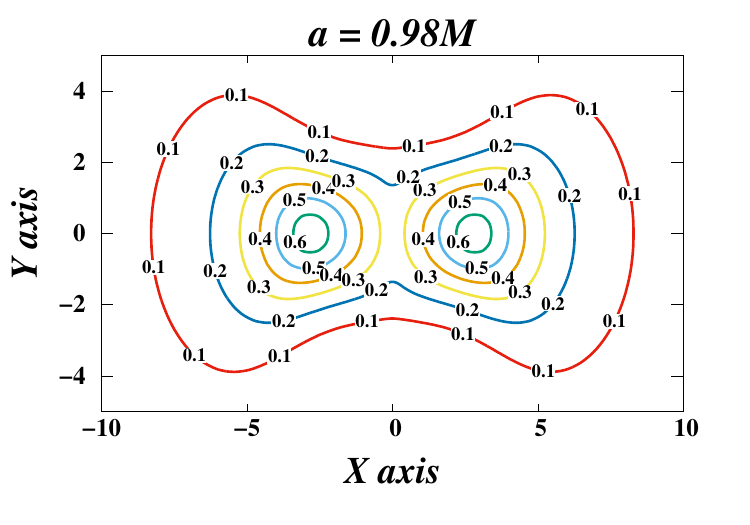}
 }
 \hspace{-5mm}
 \subfigure[]
 {
   \includegraphics[width=0.49\hsize]{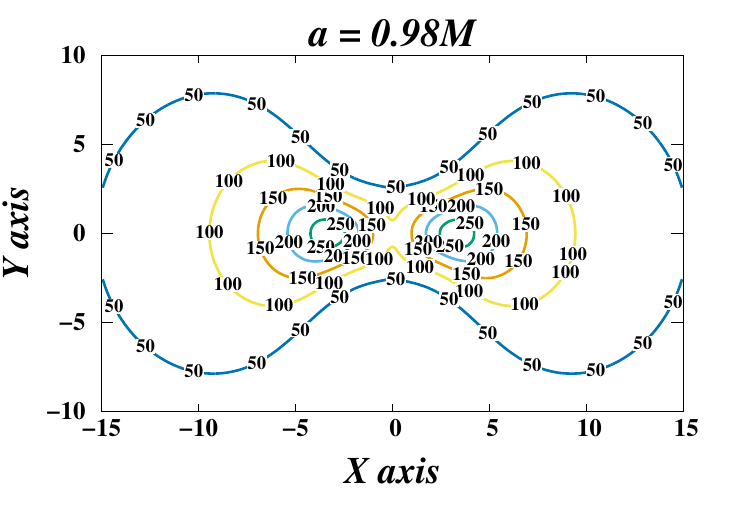}
 }
\protect 
\caption{Distribution profiles for the co-rotating disk.
  (a), (c) and (e): The distributions of the normalized electron
  number density, $n_{e}/10^{18}\,\text{cm}^{-3}$ for the BH spin
  $a\,=\,2\times 10^{-2} M, 0.50 M$ and $0.98 M$, respectively.
  (b), (d), (f): The distributions of the effective toroidal
  magnetic field, $|\mathbf{B}|^{(\mathrm{tor})}$, for the BH spin
  $a\,=\,2\times 10^{-2} M, 0.50 M$ and $0.98 M$, respectively.
  The distances in the horizontal and vertical axes are
  in units of $r_g$.}
\label{fig:dist_co_rot_dens_mag}
\end{figure}

\begin{figure}[htbp]
\centering
\subfigure[]
 {
   \includegraphics[width=0.49\hsize]{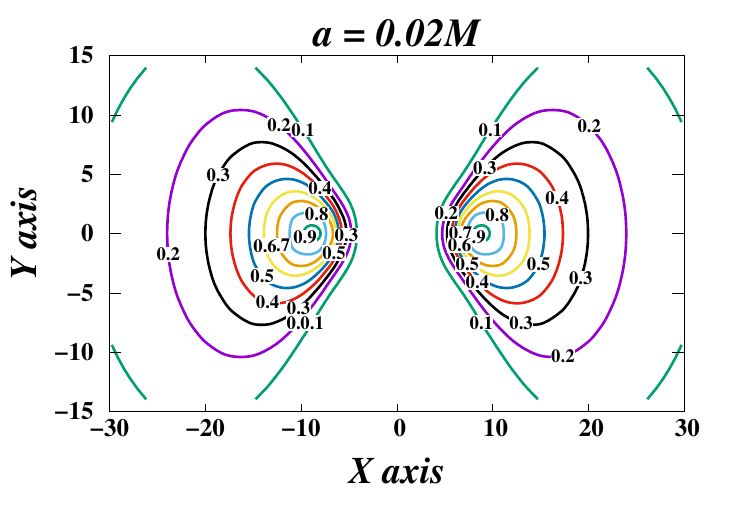}
 }
 \hspace{-5mm}
 \subfigure[]
 {
   \includegraphics[width=0.49\hsize]{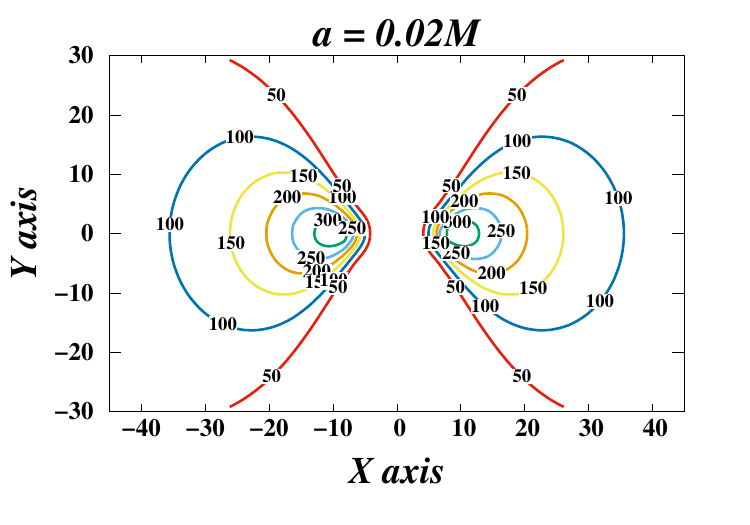}
 }
\subfigure[]
 {
   \includegraphics[width=0.49\hsize]{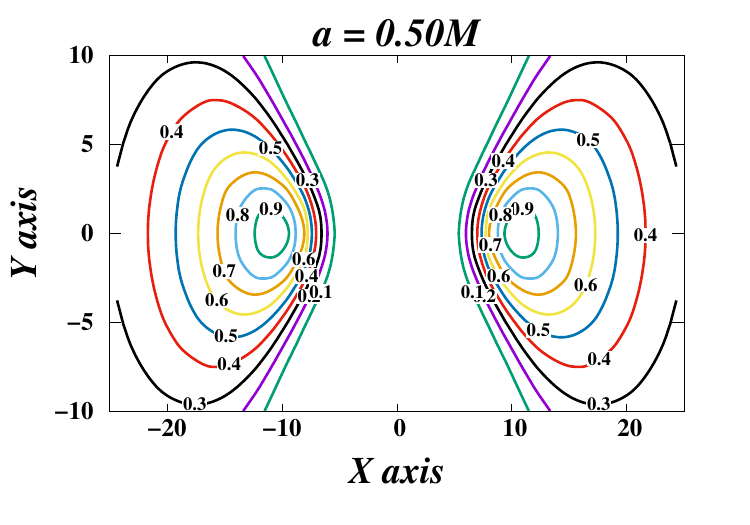}
 }
 \hspace{-5mm}
 \subfigure[]
 {
   \includegraphics[width=0.49\hsize]{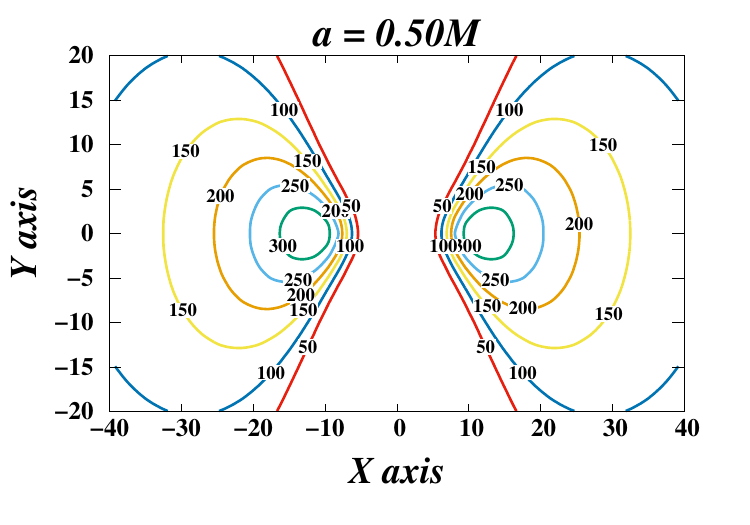}
 }
\subfigure[]
 {
   \includegraphics[width=0.49\hsize]{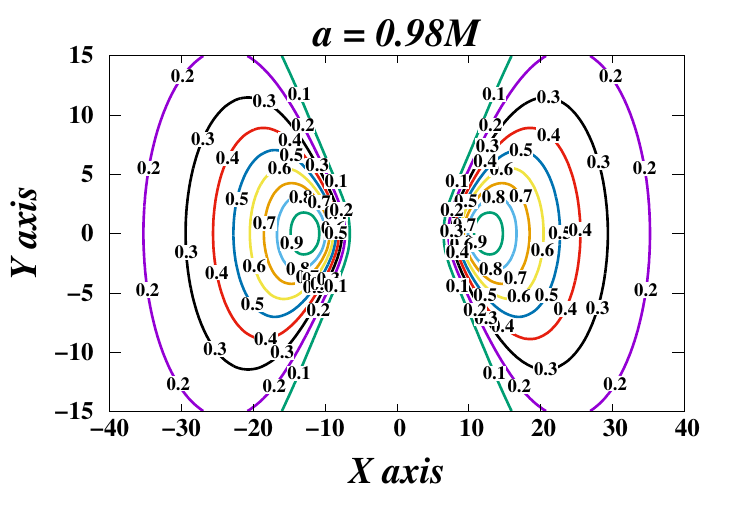}
 }
 \hspace{-5mm}
 \subfigure[]
 {
   \includegraphics[width=0.49\hsize]{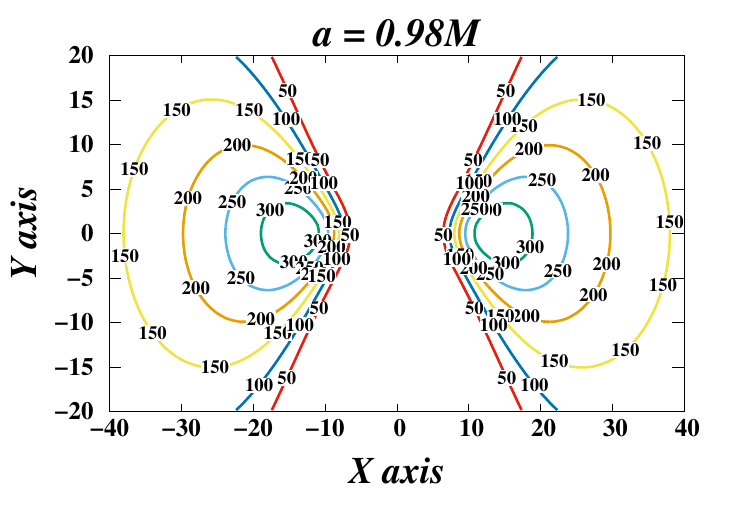}
 }
\protect 
\caption{Distribution profiles for the counter-rotating disk.
  (a), (c) and (e): The distributions of the normalized electron
  number density, $n_{e}/10^{18}\,\text{cm}^{-3}$ for the BH spin
  $a\,=\,2\times 10^{-2} M, 0.50 M$ and $0.98 M$, respectively.
  (b), (d), (f): The distributions of the effective toroidal
  magnetic field, $|\mathbf{B}|^{(\mathrm{tor})}$, for the BH spin
  $a\,=\,2\times 10^{-2} M, 0.50 M$ and $0.98 M$, respectively.
  The distances in the horizontal and vertical axes are in
  units of $r_g$.}
\label{fig:dist_counter_rot_dens_mag}
\end{figure}

In order to determine the parameters $\tilde{K}$ and $\tilde{K}_{m}$, we first choose the  maximal
number density of electrons to be
$n_e^{\mathrm{max}} = 10^{18}\,\text{cm}^{-3}$~\cite{Jia19} such that the
normalized electron number density distributions is $n_{e}/10^{18}\,\text{cm}^{-3}$,
where $n_{e}=\rho/m_{p}$, $m_p$ being the mass of proton. This provides us
the value of $\rho^{\mathrm{max}}$. For the maximal strength of the toroidal field,
we set it to be at $|\mathbf{B}|^{(\mathrm{tor})}_{\mathrm{max}} = 320\,\text{G}$.
This is $\sim 1\%$ of the Eddington limit for this type of SMBH~\cite{Bes10}.
Using Eq.~\eqref{eq:rhopm}, we simultaneously vary $\tilde{K}$ and $\tilde{K}_{m}$
to match to $\rho^{\mathrm{max}}$ and
$|\mathbf{B}|^{(\mathrm{tor})}_{\mathrm{max}} =\sqrt{2p_{m}^{\mathrm{(tor)}}}$
for each spin of BH. Note that the value of the generating
function, $W$, also depends whether the disk is co-rotating or
counter-rotating. Therefore, one has to find the values of
$\tilde{K}$ and $\tilde{K}_{m}$ for both the co-rotating and 
counter-rotating disks for each BH spin. The distributions of
the electron number density and  effective toroidal field,
$|\mathbf{B}|^{(\mathrm{tor})}$, for both the co-rotating and 
counter-rotating disks are shown in Figs.~\ref{fig:dist_co_rot_dens_mag}
and~\ref{fig:dist_counter_rot_dens_mag}. The values of $\tilde{K}$ and
$\tilde{K}_{m}$  are listed in Table~\ref{tab:num_para_2}.

For our numerical work, we use more than $288$ cores of IceLake
and $1920$ SkyLake processors in Govorun super-cluster at the Joint Institute for Nuclear Research. The number of gravitationally
scattered neutrinos we use for each combination of BH spin and
incident angle is more than $2$ million. The detailed numbers
are listed in Table~\ref{tab:number_of_neutrino}.

Here we would like to emphasize that unlike in the  previous
studies~\cite{Dvo23c,Dvo23d,Dvo23a,Dvo23b,Deka:2023ljj} where \texttt{MATLAB}
based codes were used, the numerical studies for this work are based on a
freshly written \texttt{C++} code which is more efficient, optimized and
flexible. Although it closely follow the old \texttt{MATLAB} code, it is a
totally independent code. In the process rewriting the code in \texttt{C++},
we have realized a significant incorrectness in the results produced by the
old \texttt{MATLAB} code when we considered only toroidal field in the
accretion disk. The values of two dimensionless coefficients in
Eqs.~\eqref{eq:omega_em_dimensionless}
and~\eqref{eq:omega_m_dimensionless},
$V_{\mathrm{B}} \sim O(10^{-76})$ and $V_m \sim O(10^{-87})$
(e.g. see Ref.~\cite{Dvo23b}),
respectively,  turn out to be quite
low
for \texttt{MATLAB}'s default precision. As a result, the \texttt{MATLAB}
code produced the numerical values of plasma density and magnetic pressure to be
zero's. In such a situation, the effect of the spin precession is not
visible if we consider a toroidal magnetic field only. This issue is ratified
  in~\cite{DekDvo25} as well as in this work, and we present our new results by
  considering only toroidal field in the accretion disk.

\begin{figure}[htbp]
\centering
    {
      \hspace*{-10mm}
   \includegraphics[width=1.15\hsize]{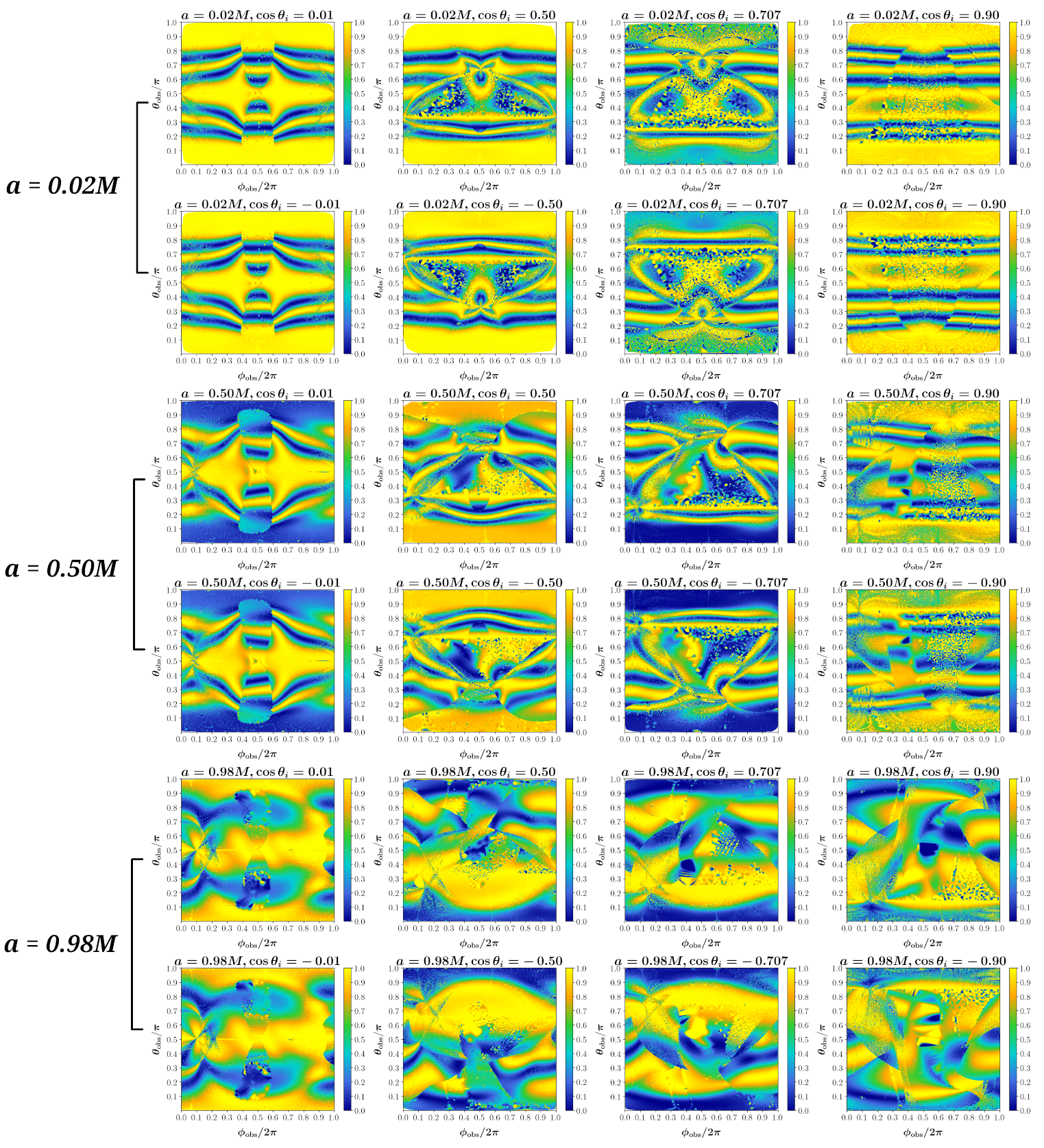}
 }
\protect 
\caption{The distributions of the survival probability, $P_{\mathrm{LL}}$ at
  $(\theta_{\mathrm{obs}}, \phi_{\mathrm{obs}})$ for the co-rotating accretion
  disks. Figures show three BH spins
  $a\,=\,2\times 10^{-2} M, 0.50 M$ and $0.98 M$ and  eight 
  incident angles, $\cos\theta_i = \pm 0.01, \pm 0.50. \pm 0.707$
  and $\pm 0.90$.
  \label{fig:corotating_all}
}
\end{figure}

\begin{figure}[htbp]
\centering
    {
      \hspace*{-10mm}
   \includegraphics[width=1.15\hsize]{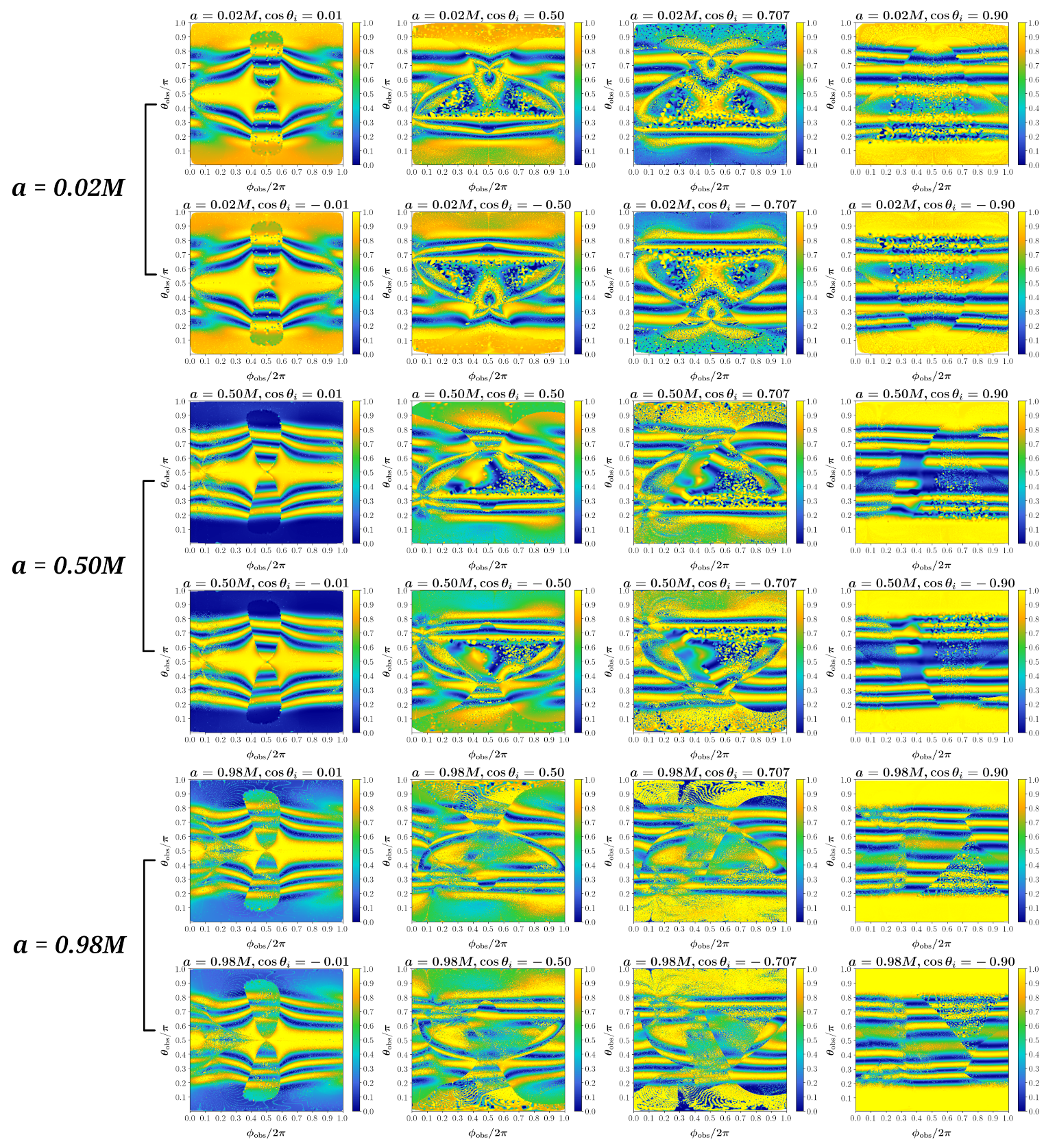}
 }
\protect 
\caption{The distributions of the survival probability, $P_{\mathrm{LL}}$ at
  $(\theta_{\mathrm{obs}}, \phi_{\mathrm{obs}})$ for the counter-rotating accretion
  disks. Figures show three BH spins
  $a\,=\,2\times 10^{-2} M, 0.50 M$ and $0.98 M$ and  eight 
  incident angles, $\cos\theta_i = \pm 0.01, \pm 0.50. \pm 0.707$
  and $\pm 0.90$.
  \label{fig:counterrotating_all}
}
\end{figure}

\section{Results and Discussion}
\label{sec:RES}

In this study, we consider three BH spins, namely
$a\,=\,2\times 10^{-2} M, 0.50 M$ and $0.98 M$. For each BH spin, we consider
eight different incident angles,
$\cos\theta_i = \pm 0.01, \pm 0.50. \pm 0.707$ and $\pm 0.90$. For each
combination of BH spin and incident angle, we consider more than $2$ million
neutrinos as shown in the Table~\ref{tab:number_of_neutrino}.

If the spin of a left-handed Dirac neutrino precesses in an external field, it
becomes sterile or right-handed. Such a neutrino cannot be observed in a detector.
Hence, the observed flux of neutrinos will be reduced by a factor of
$P_{\mathrm{LL}}$ in comparison to the flux of non-spinning particles.

Before we discuss the contribution of the external fields in an accretion
disk to the neutrino spin oscillations, we make a remark on the gravity
induced neutrino spin precession. Suppose that an ultrarelativistic neutrino
is gravitationally scattered by a BH without an accretion disk. In this case,
the spin flip has been found in Ref.~\cite{DolDorLas06} to be absent
if one deals with a nonrotating BH. This result has been confirmed in
Ref.~\cite{Dvo20}. If we study this kind of scattering by a rotating BH, the
absence of the fermion spin-flip has been established in Ref.~\cite{Far16} with
the help of geometrical methods in case of a the particle motion in the
equatorial plane. We obtained the same result in Ref.~\cite{Dvo21}. To the best
of our knowledge, there is no general proof that $P_{\mathrm{LL}} = 1$ for an
ultrarelativistic fermion scattered by a rotating BH along an arbitrary trajectory
if only the gravitational interaction is accounted for (see also Sec.~\ref{sec:SPIN_EV}). Nevertheless, we check this
statement numerically for any test particles involved in our simulations.
We get that $P_{\mathrm{LL}} = 1$ is fulfilled with high level of accuracy.

We present our results for the co-rotating disk in Fig.~\ref{fig:corotating_all}
and for the counter-rotating disk in Fig.~\ref{fig:counterrotating_all} 
as functions of $\theta_{\mathrm{obs}}$ and $\phi_{\mathrm{obs}}$. The angles
at which the incoming neutrinos are incident are mentioned above. Also as
mentioned earlier, we consider only toroidal magnetic field in the accretion disk.
All the areas with $P_{\mathrm{LL}} < 1$ signify spin flip
of neutrinos. The lower the value of $P_{\mathrm{LL}}$ is, the higher the probability
of the spin flip is. We see that in all cases in Figs.~\ref{fig:corotating_all}
and~\ref{fig:counterrotating_all}, there are non-negligible probabilities of spin flip.

The result we obtain is different from what is reported in the
studies~\cite{Dvo23c,Dvo23d,Dvo23a,Dvo23b,Deka:2023ljj}. In particular, we find 
that a sizable neutrino spin-flip takes place even we consider the presence of
only toroidal magnetic field in contrary to
Refs.~\cite{Dvo23c,Dvo23d,Dvo23a,Dvo23b,Deka:2023ljj} where a poloidal component
is considered. As mentioned in Sec.~\ref{sec:NUMERICAL}, this is due to the
precision related issue involved with
the \texttt{MATLAB} code that has been used in those studies. The default numerical
precision of \texttt{MATLAB} was unable to deal with the smallness of the
two dimensionless coefficients, $V_m$ and $V_{\mathrm{B}}$. We have corrected this
issue in our new \texttt{C++} code.

A neutrino can undergo spin oscillations when it interacts with a magnetic field
transverse to its velocity. One can see in Fig.~\ref{fig:tourus_4} that there exist
neutrinos which cross the accretion disk while the toroidal magnetic field is
perpendicular to their momenta. Using such a qualitative argument, we conclude that only toroidal field can cause the neutrino spin-flip in the gravitational scattering.

This finding of ours is promising since the toroidal magnetic field is inherent in the ``Polish doughnut'' model. Despite the fact that the helical magnetic fields, i.e. having both toroidal and poloidal components, provide the stabilization of astrophysical plasmas~\cite{CanSor23}, there is no commonly accepted model for a poloidal field in accretion disks.

The next important improvement made in the present work in comparison to Refs.~\cite{Dvo23c,Dvo23b} consists in the explicit separate descriptions of spin oscillations of upper and lower particles. Indeed, in Refs.~\cite{Dvo23c,Dvo23b}, we have computed the survival probability only for the upper particles and then reconstructed $P_\mathrm{LL}$ for the lower ones using the symmetry properties of the effective Hamiltonian in Eq.~\eqref{eq:effHam}. It leads to the symmetric distributions of $P_\mathrm{LL}(\theta_\mathrm{obs},\phi_\mathrm{obs})$ with respect to the $\theta_\mathrm{obs}=\pi/2$ reflection plane.

However, such a reflection symmetry with respect to the $\theta_\mathrm{obs}=\pi/2$ plane is valid only for purely gravitational interactions of neutrinos in the absence of the accretion disk. If one takes into account the neutrino interactions with the background matter and magnetic field, $P_\mathrm{LL}(\theta_\mathrm{obs},\phi_\mathrm{obs})$ remains symmetric only for a slowly rotating BH, $a\ll 1$, while
the incoming neutrinos propagate parallel to the equatorial plane, $|\cos\theta_i| \ll 1$. In all other cases, $P_\mathrm{LL}(\theta_\mathrm{obs},\phi_\mathrm{obs})$ is not symmetric with respect to the $\theta_\mathrm{obs}=\pi/2$ plane. This feature can be seen in Figs.~\ref{fig:corotating_all} and~\ref{fig:counterrotating_all}.

In the present work, we describe spin oscillations of neutrinos for the arbitrary inclination of the incoming flux, $0 < \theta_i < \pi$. One can see in Figs.~\ref{fig:corotating_all} and~\ref{fig:counterrotating_all} that $P_\mathrm{LL}$ for the opposite values of $\cos\theta_i$ with the same BH spin are inversely symmetric only in the case of a slowly rotating BH, $a\ll 1$. This feature of $P_\mathrm{LL}$ disappears for a rapidly rotating BH. It means that it is an inherent property of the Kerr metric.

The study with a number of arbitrary $\theta_i$'s is motivated by the fact that, in some cases, the BH spin can be significantly inclined towards the galactic plane. For example, there are indirect indications in Ref.~\cite{Aki22V} that the spin of SMBH in our Galaxy is almost in the Galactic plane. If we study the gravitational scattering and spin oscillations of neutrinos, which are emitted in a SN
explosion, by such a BH, our results shall be of great importance. Since we do not know the mutual position of a neutrino source and the Earth at the moment of the observation of the neutrino signal, one has to consider multiple values of $\theta_i$. In our work, we attempt
  to address this issue.

The next important result obtained in the present work is the consideration of both co-rotating and counter-rotating disks. Comparing Figs.~\ref{fig:corotating_all} and~\ref{fig:counterrotating_all}, one can see that the observed fluxes of scattered neutrinos are different despite we consider the disks with equal maximal number densities of plasma and maximal strengths of the magnetic field. Thus, performing a precise measurement of astrophysical neutrinos
that are gravitationally scattered by SMBH, one can potentially infer the information about the disk rotation direction.

\section{Conclusion}
\label{sec:CONCL}

In this work, we consider the propagation of the
ultra-relativistic neutrinos with
nonzero magnetic moment in the strong gravitational field of a
SMBH. The neutrinos are traveling at a random angle with respect
to the equatorial plane of the SMBH. The SMBH is surrounded by
a thick magnetized accretion disk. We consider those neutrinos
that are scattered due to strong gravitational field of the SMBH
since only such neutrinos can be observed. We can exactly
describe the geodesic motion of these neutrinos in Kerr metric.

The neutrinos not only interact electroweakly with the rotating
matter of the accretion disk, but also interact magnetically with
the disk due to its nonzero magnetic moment. In this study, we
consider only toroidal field inside the disk. The neutrino
interaction with the external fields causes the spin precession
which can be accounted for along each neutrino trajectory. To
have a good resolution of $P_{\mathrm{LL}}$ in the
$(\theta_{\mathrm{obs}}, \phi_{\mathrm{obs}})$ plane, we consider
more than $2$ million gravitationally scattered neutrinos in
each combination of BH spin and incident angle.

The importance of our results is multi-fold. First,
Figs.~\ref{fig:corotating_all} and~\ref{fig:counterrotating_all}
show that it is enough to have the toroidal magnetic field, which is
inherent in the ``Polish doughnut'' model, for spin oscillations
to occur.

Second, we find that the symmetric distributions of
$P_\mathrm{LL}(\theta_\mathrm{obs},\phi_\mathrm{obs})$ with respect
to the $\theta_\mathrm{obs}=\pi/2$ reflection plane can be seen
either in the absence of the accretion disk or for 
a slowly rotating BH in the presence of the accretion disk.
For all other cases, no such symmetry is observed.

Third, we observe that the inverse symmetry of $P_\mathrm{LL}$
for the opposite values of $\cos\theta_i$ with the same BH spin
is exhibited for a slowly rotating BH only. For a rapidly
rotating BH, this inverse symmetry disappears. This
points to the inherent property of the Kerr metric.

Fourth, since the relative position between a neutrino source
and  Earth is not known during the observation, our study
carries important information due to the computation of
$P_\mathrm{LL}$  for a
number of arbitrary $\theta_i$'s.

Fifth, our results show the clear difference between the
observed fluxes of scattered neutrinos for the  co-rotating and
counter-rotating disks (see Figs.~\ref{fig:corotating_all}
and~\ref{fig:counterrotating_all}). This can
help in inferring the direction of the disk rotation in a precise
measurement of astrophysical neutrinos that are gravitationally
scattered by SMBH.

One can see in Figs.~\ref{fig:corotating_all} and~\ref{fig:counterrotating_all} that the neutrino spin-flip can be quite sizable for certain outgoing neutrinos. The particles with small $P_\mathrm{LL}$ are scattered to the areas marked by the blue color in Figs.~\ref{fig:corotating_all} and~\ref{fig:counterrotating_all}. It means that an observer will detect almost no neutrino signal if placed in these positions. We mention that such a significant spin-flip is reached in quite weak magnetic fields $B \lesssim 3\times10^2\,\text{G}$ and for neutrinos having a moderate magnetic moment $10^{-13}\mu_\mathrm{B}$. This result can be explained by the fact that we consider the neutrino propagation near a SMBH. The length parameter in our situation is $r_g = 2\times10^8 M_\odot = 3\times10^{13}\,\text{cm}$. Thus, if one normalizes tiny magnetic and electroweak potentials to $r_g$, the resulting dimensionless quantities are not small.

The results of the present work can be used for the studies of
the distributions of magnetic fields and background matter in the
vicinity of SMBH in the center of our Galaxy. Suppose that a core
collapsing SN is exploded somewhere in the Galaxy. Existing and future
neutrino telescopes are able to detect up to $\sim 10^4$ SN neutrinos
which propagate directly to the Earth (see, e.g.,  Ref.~\cite{Abu22}).
It allows one to calibrate the neutrino signal. Note that the Galactic
center is not necessarily on the SN-Earth line. Then, one should try
to detect the same SN neutrinos lensed by the SMBH in the center of
the Galaxy. Of course, there will be much less number of such neutrinos
in comparison with the direct flux from SN. These particles are affected
by the strong gravity, with their spin being precessed in external fields
near SMBH. Such an observation will potentially allow one to study the
magnetic fields in the Galactic center by their influence on neutrino
spin oscillations.


\section*{Acknowledgments}
  We thank Yu.~N.~Obukhov, O.~V.~Teryaev, and A.~F.~Zakharov for the useful discussions. All our
  numerical computations have been performed at Govorun
  super-cluster at Joint Institute for Nuclear Research, Dubna.

\appendix

\section{Explicit forms for various quantities}
\label{app:explicit_form}

\subsection{The components of gravi-electromagnetic field}
\label{app:sub_gravi_electro_mag_field}

The explicit form of $\tilde{\bm{e}}_{g}$
and $\tilde{\bm{b}}_{g}$ in
Eq.~\eqref{eq:omega_gravito_magnetic_dimensionless} is given in~\cite{Dvo23a}. These vectors are
%
\begin{align}\label{eq:egbg}
  \tilde{e}_{g1} = & \frac{1}
  {2\sqrt{z^{2}\cos^{2}\theta(x^{2}-x+z^{2})+z^{2}x^{2}+z^{2}x+x^{4}}(x^{2}+z^{2}\cos^{2}\theta)^{2}}
  \\
  & \times
  \Big\{
    r_{g}\frac{\mathrm{d}\phi}{\mathrm{d}r}
    \left[
      z^{3}\cos^{4}\theta(z^{2}-x^{2})-z\cos^{2}\theta(z^{4}+3x^{4})+z^{3}x^{2}+3zx^{4}
    \right]
    \\
    & +
    \frac{\mathrm{d}t}{\mathrm{d}r}
    \left[
      z^{2}\cos^{2}\theta(z^{2}+x^{2})-z^{2}x^{2}-x^{4}
    \right]
  \Big\},
  \nonumber
  \displaybreak[2]
  \\
  \tilde{e}_{g2} = & \frac{z^{2}x\sin2\theta\sqrt{x^{2}-x+z^{2}}
  \left[
    \frac{\mathrm{d}t}{\mathrm{d}r}-r_{g}\frac{\mathrm{d}\phi}{\mathrm{d}r}z\sin^{2}\theta
  \right]}
  {2\sqrt{z^{2}\cos^{2}\theta(x^{2}-x+z^{2})+z^{2}x^{2}+z^{2}x+x^{4}}(x^{2}+z^{2}\cos^{2}\theta)^{2}},
  \nonumber
  \displaybreak[2]
  \\
  \tilde{e}_{g3} = & -\frac{
  \left[
    z^{2}x\sin2\theta r_{g}\frac{\mathrm{d}\theta}{\mathrm{d}r}(x^{2}-x+z^{2})+z^{2}\cos^{2}\theta(z^{2}-x^{2})-z^{2}x^{2}-3x^{4}
  \right]}
  {2
  \left[
      z^{4}\cos^{4}\theta(x^{2}-x+z^{2})+xz^{2}\cos^{2}\theta(2z^{2}x+2x^{3}-x^{2}+z^{2})+x^{6}+z^{2}x^{4}+z^{2}x^{3}
  \right]}
  \\
  & \times
  \frac{z\sin\theta}{\sqrt{x^{2}-x+z^{2}}},
  \nonumber
  \displaybreak[2]
  \\
  \tilde{b}_{g1} = & \frac{\cos\theta}
  {\sqrt{z^{2}\cos^{2}\theta(x^{2}-x+z^{2})+z^{2}x^{2}+z^{2}x+x^{4}}(x^{2}+z^{2}\cos^{2}\theta)^{2}}
  \\
  & \times
  \Big\{
    r_{g}
    \frac{\mathrm{d}\phi}{\mathrm{d}r}
    \left[
      z^{4}\cos^{4}\theta(x^{2}-x+z^{2})+2x^{2}z^{2}\cos^{2}\theta(x^{2}-x+z^{2})+z^{4}x+z^{2}x^{4}+2z^{2}x^{3}+x^{6}
    \right]
    \displaybreak[2]
    \\
    & -
    zx\frac{\mathrm{d}t}{\mathrm{d}r}(z^{2}+x^{2})
  \Big\},
  \nonumber
  \displaybreak[2]
  \\
  \tilde{b}_{g2} = & -\frac{\sin\theta\sqrt{x^{2}-x+z^{2}}}
  {2\sqrt{z^{2}\cos^{2}\theta(x^{2}-x+z^{2})+z^{2}x^{2}+z^{2}x+x^{4}}(x^{2}+z^{2}\cos^{2}\theta)^{2}}
  \displaybreak[2]
  \\
  & \times
  \left\{
    r_{g}\frac{\mathrm{d}\phi}{\mathrm{d}r}
    \left[
      z^{4}\cos^{4}\theta(2x-1)+z^{2}\cos^{2}\theta(z^{2}+4x^{3}+x^{2})+2x^{5}-z^{2}x^{2}
    \right]
%
   +
    z\frac{\mathrm{d}t}{\mathrm{d}r}(x^{2}-z^{2}\cos^{2}\theta)
  \right\},
  \nonumber
  \displaybreak[2]
  \\
  \tilde{b}_{g3} = & \frac{z^{2}\sin\theta\cos\theta+r_{g}\frac{\mathrm{d}\theta}{\mathrm{d}r}x(x^{2}-x+z^{2})}
  {\sqrt{x^{2}-x+z^{2}}(x^{2}+z^{2}\cos^{2}\theta)}.
\end{align}
%
In Eq.~\eqref{eq:egbg}, the derivatives such as,
$\frac{\mathrm{d}\phi}{\mathrm{d}r}$ etc., can be calculated with the help of
Eqs.~\eqref{eq:trajth_1} and~\eqref{eq:trajphi_1}.
However, we have to take into account the number of inversions of the trajectory.
Thus, we get,
%
\begin{align}\label{eq:trajderiv}
  r_{g}\frac{\mathrm{d}\theta}{\mathrm{d}r} = &
  (-1)^{N}\frac{\sqrt{w-(y^{2}+w-z^{2})\cos^{2}\theta-z^{2}\cos^{4}\theta}}{\sin\theta\sqrt{x^{4}+x^{2}
  \left[
    z^{2}-w-y^{2}
  \right]+x
  \left[
    w+
    \left(
      z-y
    \right)^{2}
  \right]
  -z^{2}w}},
  \displaybreak[2]
  \\
  r_{g}\frac{\mathrm{d}\phi}{\mathrm{d}r} = &
  \pm\frac{z\sin^{2}\theta(x-zy)+y(x^{2}-x+z^{2})}{\sin^{2}\theta(x^{2}-x+z^{2})\sqrt{x^{4}+x^{2}
  \left[
    z^{2}-w-y^{2}\right]+x
    \left[
      w+
      \left(
        z-y
      \right)^{2}
  \right]
  -z^{2}w}},
  \displaybreak[2]
  \\
  \frac{\mathrm{d}t}{\mathrm{d}r}= &
  \pm\frac{z^{2}\cos^{2}\theta(x^{2}-x+z^{2})+z^{2}x^{2}+z^{2}x+x^{4}-xyz}{(x^{2}+z^{2}-x)\sqrt{x^{4}+x^{2}
  \left[
    z^{2}-w-y^{2}
  \right]
  +x
  \left[
    w+
    \left(
      z-y
    \right)^{2}
  \right]
  -z^{2}w}},
\end{align}
%
where the signs $\pm$ correspond to outgoing and incoming particles, and the
number of inversions, $N$, is given in Eqs.~\eqref{eq:Ninv_in_north},
\eqref{eq:Ninv_out_north}, \eqref{eq:Ninv_in_south} and~\eqref{eq:Ninv_out_south}.

\subsection{The effective velocity}
\label{app:sub_minkowskian_vel}

The components of the vector $\bm{v}$, used in Eq.~\eqref{eq:omega_gravito_magnetic_dimensionless}, have the form,
%
\begin{align}\label{eq:trajv}
  v_{1} = &
  \pm\sqrt{x^{4}+x^{2}\left[z^{2}-w-y^{2}\right]+x\left[w+\left(z-y\right)^{2}\right]-z^{2}w}
  \displaybreak[2]
  \\
  & \times
  \frac{\sqrt{z^{2}\cos^{2}\theta(x^{2}-x+z^{2})+z^{2}x^{2}+z^{2}x+x^{4}}}{z^{2}\cos^{2}\theta(x^{2}-x+z^{2})+z^{2}x^{2}+z^{2}x+x^{4}-xyz},
  \\
  v_{2} = &
  \pm(-1)^{N}\sqrt{w-(y^{2}+w-z^{2})\cos^{2}\theta-z^{2}\cos^{4}\theta}
  \displaybreak[2]
  \\
  & \times
  \frac{\sqrt{z^{2}+x^{2}-x}\sqrt{z^{2}\cos^{2}\theta(x^{2}-x+z^{2})+z^{2}x^{2}+z^{2}x+x^{4}}}
  {\sin\theta\left[z^{2}\cos^{2}\theta(x^{2}-x+z^{2})+z^{2}x^{2}+z^{2}x+x^{4}-xyz\right]},
  \displaybreak[2]
  \\
  v_{3} = & \frac{y\left\{ (x^{2}+z^{2}\cos^{2}\theta-x)\left[z^{2}\cos^{2}\theta(x^{2}-x+z^{2})+z^{2}x^{2}+z^{2}x+x^{4}\right]+x^{2}z^{2}\sin^{2}\theta\right\} }{\sin\theta(x^{2}+z^{2}\cos^{2}\theta)\sqrt{x^{2}-x+z^{2}}\left[z^{2}\cos^{2}\theta(x^{2}-x+z^{2})+z^{2}x^{2}+z^{2}x+x^{4}-xyz\right]}.
\end{align}
%
The quantity $u^0$ reads
\begin{equation}\label{eq:u0}
  \frac{u^{0}}{U^{t}} =\frac{\sqrt{z^{2}\cos^{2}\theta+x^{2}}\sqrt{z^{2}+x^{2}-x}}{\sqrt{z^{2}\cos^{2}\theta(x^{2}-x+z^{2})+z^{2}x^{2}+z^{2}x+x^{4}}}.
\end{equation}
We do not provide the components of the four vector $l^a$,
$l^{0}=\frac{\mathrm{d}t}{\mathrm{d}r}\frac{u^{0}}{U^{t}}$
and $\mathbf{l}=\frac{\mathrm{d}t}{\mathrm{d}r}\frac{u^{0}}{U^{t}}\mathbf{v}$ here,
since they can be computed based on Eqs.~\eqref{eq:trajderiv}-\eqref{eq:u0}.

\subsection{The effective potential of the neutrino interaction with background matter}
\label{app:sub_minkowskian_backgnd_potential}

From Eq.~\eqref{eq:Gmu_3}, we can write the nonzero
  components of the four vector $g^a = (g^0,0,0,g^3)$ as,
%
\begin{align}\label{eq:gcomp}
  g^{0} & =\frac{\sqrt{x^{2}+z^{2}\cos^{2}\theta}
  \sqrt{x^{2}-x+z^{2}}U_{f}^{t}}
  {\sqrt{z^{2}\cos^{2}\theta(x^{2}-x+z^{2})+z^{2}x^{2}+z^{2}x+x^{4}}},  
  \nonumber
  \displaybreak[2]
  \\
  g^{3} & =\frac{\sin\theta
  \left[
    r_{g}U_{f}^{\phi}
    \left(
      z^{2}\cos^{2}\theta(x^{2}-x+z^{2})+z^{2}x^{2}+z^{2}x+x^{4}
    \right)-
    U_{f}^{t}xz
  \right]}
  {\sqrt{x^{2}+z^{2}\cos^{2}\theta}\sqrt{z^{2}\cos^{2}\theta
  (x^{2}-x+z^{2})+z^{2}x^{2}+z^{2}x+x^{4}}}.
\end{align}

We use the above forms of $g^0$ and $g^3$ in
  Eq.~\eqref{eq:omega_m_dimensionless} to compute
  $\bm{\bm{\Omega}}_{x}^{\mathrm{matt}}$.

\subsection{Toroidal magnetic field in the locally Minkowskian frame}
\label{app:sub_minkowskian_toroidal_field}

The only nonzero component of the electromagnetic field in the locally
Minkowskian frame, which contributes to Eq.~\eqref{eq:omega_em_dimensionless},
is
%
\begin{align}\label{eq:bztor}
  b_{3}= & -\frac{U_{f}^{t}r_{g}^{2}\sqrt{2p_{m}}}
  {\sin\theta(1-\Omega l_{0})
  \sqrt{(x^{2}+z^{2}\cos^{2}\theta)(x^{2}-x+z^{2})}}
  \nonumber
  \\
  & \times
  \big\{
    \lambda_{0}^{2}(x^{2}-x+z^{2}\cos^{2}\theta)+
    2\lambda_{0}xz\sin^{2}\theta
    - \sin^{2}\theta
    \left[
      (x^{2}+z^{2})(x^{2}+z^{2}\cos^{2}\theta)+xz^{2}\sin^{2}\theta
    \right]
  \big\}^{1/2},
\end{align}
%
where $\lambda_{0}=l_{0}/r_{g}$.

\section{Elliptic integrals and their useful relations}
\label{app:elliptic_integrals}

We follow the definitions of Elliptic integrals and functions as in
Ref.~\cite{AbrSte64}. In this Appendix, we provide some of the useful expressions.

The Incomplete Elliptic integral of first kind is defined as
\eqarray{\label{eq:F}
  F(\varphi, m) &=& \displaystyle\int^\varphi_0 \frac{\mathrm{d}x}{\sqrt{1 - m \sin^2 x}}.
}
If $\varphi = \frac{\pi}{2}$, we get the complete Elliptic integral of first
kind in the form,
\eqarray{\label{eq:K}
  K(m) &\equiv& F\left(\frac{\pi}{2}, m\right) \,=\, \displaystyle\int^{\pi/2}_0 \frac{\mathrm{d}x}{\sqrt{1 - m \sin^2 x}}.
}

The Incomplete Elliptic integral of third kind is defined as,
\eqarray{
  \Pi(\varphi, n, m) &=&
  \displaystyle\int^\varphi_0 \frac{\mathrm{d}x}{(1 - n \sin^2 x) \sqrt{1 - m \sin^2 x}}.
}
If $\varphi = \frac{\pi}{2}$, we get the complete Elliptic integral of third
kind,
\eqarray{
  \Pi(n, m) &\equiv& \Pi\left(\frac{\pi}{2}, n, m\right) \,=\,
  \displaystyle\int^{\pi/2}_0 \frac{\mathrm{d}x}{(1 - n \sin^2 x) \sqrt{1 - m \sin^2 x}}.
}

The functions $F(\varphi, m)$ and $K(m)$ obey the following properties:
\eqarray{
  F(\varphi, - m) &=& \frac{1}{\sqrt{1+m}}
  \left[
    K \left(\frac{m}{1+m}\right) - F \left(\frac{\pi}{2} - \varphi, \frac{m}{1+m}\right)
    \right],\nonumber\\
  K (-m) &=& \frac{1}{\sqrt{1+m}} K \left(\frac{m}{1+m}\right),
  \label{eq:elliptic_integrals_1}
}
which can be derived from Eqs.~\eqref{eq:F} and~\eqref{eq:K}.

\section{An example of the calculation of the integrals\label{sec:INTCALC}}

In this appendix, we provide the detailed computation of one of the $\theta$-integrals resulting from Eq.~\eqref{eq:trajth_1}.

Let us consider the motion of the upper particles before the turn point. The variable $\mathfrak{\tilde t} = \cos\theta$ changes from $\mathfrak{\tilde t}_i$ to the current value $\mathfrak{\tilde t}$. However, it reaches $\mathfrak{\tilde t}$ oscillating $N^\mathrm{before}$ times between the borders
$\pm \mathfrak{\tilde t}_+$. Let us take that $N^\mathrm{before}$ is odd and $N^\mathrm{before} \geq 1$.

We can split the whole range of the $\mathfrak{\tilde t}$ variation into the intervals,
\begin{equation}\label{eq:interval}
  \mathfrak{\tilde t} \in [\mathfrak{\tilde t}_i,+\mathfrak{\tilde t}_+] \cup \underbrace{ [+ \mathfrak{\tilde t}_+,-\mathfrak{\tilde t}_+] \cup \dots \cup [- \mathfrak{\tilde t}_+,+\mathfrak{\tilde t}_+] }_{(N^\mathrm{before}-1)\,\text{times}} \cup [+\mathfrak{\tilde t}_+,\mathfrak{\tilde t}].
\end{equation}
Since $\mathfrak{\tilde t}_i < \mathfrak{\tilde t}_+$ by definition, and we consider upper particles only, $\mathfrak{\tilde t}$ is ascending in the first segment in Eq.~\eqref{eq:interval}. Hence, the first $\theta$-integral has the factor of $+1$. Since $N^\mathrm{before}$ is odd and $N^\mathrm{before} \geq 1$, $\mathfrak{\tilde t}$ is descending in the last segment in Eq.~\eqref{eq:interval}. Hence, the last $\theta$-integral acquires the factor of $-1$.

Thus, the $\theta$-integral in Eq.~\eqref{eq:trajth_1}, corresponding to the partition in Eq.~\eqref{eq:interval}, reads
\begin{align}\label{eq:intsplit}
  \int_{\mathfrak{\tilde t}_{i}}^{\mathfrak{\tilde t}}(\cdots) = &
  \int_{\mathfrak{\tilde t}_{i}}^{\mathfrak{\tilde t}_+} (\cdots) +
  \underbrace{ 
  \left[
    - \int_{+\mathfrak{\tilde t}_+}^{-\mathfrak{\tilde t}_+} (\cdots) + \int_{-\mathfrak{\tilde t}_+}^{+\mathfrak{\tilde t}_+} (\cdots) - \cdots - \int_{+\mathfrak{\tilde t}_+}^{-\mathfrak{\tilde t}_+} (\cdots)
  \right]
  }_{(N^\mathrm{before}-1)\,\text{times}}
  - \int_{+\mathfrak{\tilde t}_+}^{\mathfrak{\tilde t}} (\cdots) 
  \notag
  \\
  & =
  I_{i}+(N^\mathrm{before}-1)I_{m}+I_{f},
\end{align}
where
\begin{align}\label{eq:Iimf}
  I_{i} = & \sqrt{\mathfrak{\tilde t}_{-}^{2}+\mathfrak{\tilde t}_{+}^{2}}\int_{\mathfrak{\tilde t}_{i}}^{\mathfrak{\tilde t}_{+}} \frac{\mathrm{d}\mathfrak{\tilde t}'}{\sqrt{(\mathfrak{\tilde t}_{-}^{2}+\mathfrak{\tilde t}'^{2})(\mathfrak{\tilde t}_{+}^{2}-\mathfrak{\tilde t}'^{2})}}
  \notag
  \\
  & =\frac{\sqrt{\mathfrak{\tilde t}_{-}^{2}+\mathfrak{\tilde t}_{+}^{2}}}{\mathfrak{\tilde t}_{-}}\left[K\left(-\frac{\mathfrak{\tilde t}_{+}^{2}}{\mathfrak{\tilde t}_{-}^{2}}\right)-F\left(\arcsin\frac{\mathfrak{\tilde t}_{i}}{\mathfrak{\tilde t}_{+}},-\frac{\mathfrak{\tilde t}_{+}^{2}}{\mathfrak{\tilde t}_{-}^{2}}\right)\right]
  =
  F\left(\arccos\frac{\mathfrak{\tilde t}_{i}}{\mathfrak{\tilde t}_{+}},\frac{\mathfrak{\tilde t}_{+}^{2}}{\mathfrak{\tilde t}_{-}^{2}+\mathfrak{\tilde t}_{+}^{2}}\right),
  \notag
  \\
  I_{m} = & \sqrt{\mathfrak{\tilde t}_{-}^{2}+\mathfrak{\tilde t}_{+}^{2}} \int_{-\mathfrak{\tilde t}_{+}}^{\mathfrak{\tilde t}_{+}}\frac{\mathrm{d}\mathfrak{\tilde t}'}{\sqrt{(\mathfrak{\tilde t}_{-}^{2}+\mathfrak{\tilde t}'^{2})(\mathfrak{\tilde t}_{+}^{2}-\mathfrak{\tilde t}'^{2})}}
  \notag
  \\
  & =
  \frac{2\sqrt{\mathfrak{\tilde t}_{-}^{2}+\mathfrak{\tilde t}_{+}^{2}}}{\mathfrak{\tilde t}_{-}}K\left(-\frac{\mathfrak{\tilde t}_{+}^{2}}{\mathfrak{\tilde t}_{-}^{2}}\right)
  =\frac{2\sqrt{\mathfrak{\tilde t}_{-}^{2}+\mathfrak{\tilde t}_{+}^{2}}}{\sqrt{\mathfrak{\tilde t}_{-}^{2}+\mathfrak{\tilde t}_{+}^{2}}}K\left(\frac{\mathfrak{\tilde t}_{+}^{2}}{\mathfrak{\tilde t}_{-}^{2}+\mathfrak{\tilde t}_{+}^{2}}\right),
  \notag
  \\
  I_{f} = & -\sqrt{\mathfrak{\tilde t}_{-}^{2}+\mathfrak{\tilde t}_{+}^{2}}\int_{\mathfrak{\tilde t}_{+}}^{\mathfrak{\tilde t}}\frac{\mathrm{d}\mathfrak{\tilde t}'}{\sqrt{(\mathfrak{\tilde t}_{-}^{2}+\mathfrak{\tilde t}'^{2})(\mathfrak{\tilde t}_{+}^{2}-\mathfrak{\tilde t}'^{2})}}
  \notag
  \\
  & =
  \frac{\sqrt{\mathfrak{\tilde t}_{-}^{2}+\mathfrak{\tilde t}_{+}^{2}}}{\mathfrak{\tilde t}_{-}}\left[K\left(-\frac{\mathfrak{\tilde t}_{+}^{2}}{\mathfrak{\tilde t}_{-}^{2}}\right)-F\left(\arcsin\frac{\mathfrak{\tilde t}}{\mathfrak{\tilde t}_{+}},-\frac{\mathfrak{\tilde t}_{+}^{2}}{\mathfrak{\tilde t}_{-}^{2}}\right)\right]
  = F\left(\arccos\frac{\mathfrak{\tilde t}}{\mathfrak{\tilde t}_{+}},\frac{\mathfrak{\tilde t}_{+}^{2}}{\mathfrak{\tilde t}_{-}^{2}+\mathfrak{\tilde t}_{+}^{2}}\right),
\end{align}
where we use the definition of elliptic integrals in Eqs.~\eqref{eq:F} and~\eqref{eq:K} as well as their properties in Eq.~\eqref{eq:elliptic_integrals_1}. The cases of even $N^\mathrm{before}$ and $N^\mathrm{before} = 0$ can be studied analogously.

Computing the $r$-integral in Eq.~\eqref{eq:IxIt_in_north} numerically and using Eq.~\eqref{eq:intsplit}, we get the equation
\begin{equation}\label{eq:Ninveq}
  I_x = I_{i}+(N^\mathrm{before}-1)I_{m}+I_{f}.
\end{equation}
Taking into account that $I_f/I_m <1$, we obtain Eq.~\eqref{eq:Ninv_in_north}. Inserting the values of $I_{i}$, $I_m$, and $I_f$ in Eq.~\eqref{eq:Iimf} to Eq.~\eqref{eq:Ninveq} and inverting $F\left( \arccos(\mathfrak{\tilde t}/\mathfrak{\tilde t}_{+}),\mathfrak{\tilde t}_{+}^{2}/[\mathfrak{\tilde t}_{-}^{2}+\mathfrak{\tilde t}_{+}^{2}] \right)$ with help of the elliptic Jacobi function, one derives Eq.~\eqref{eq:thetabtp_north}.

The motion of a test particle after the turn point can be described if we take the final $\mathfrak{\tilde t}$ in the above as the initial value. The expressions for the polar angles in Eqs.~\eqref{eq:phibtp_north} and \eqref{eq:phiatp_north}, as well as the description of the lower particles in Sec.~\ref{subsec:south_particles}, can be carried out analogously.

\section{Adam-Bashforth and Adam-Moulton Multi-step methods}
\label{app:ab_and_am_methods}

We discuss the solution of a first order differential equation
\eqarray{
  \frac{\mathrm{d}y}{\mathrm{d}x} &=& f(y(x),x).
}
The function $f$ is known only at discrete and irregularly spaced points, $i$.
Due to this, we use $4$th order Adam-Bashforth predictor method for approximate solution, and
then we use $4$th order Adam-Moulton corrector method to improve the obtained solution.

For a $4$th order solution, we can write for Adam-Bashforth method as,
\eqarray{
  \label{eq:adam_bashforth}
  y_{i+4}
  &=& y_{i+3}
  + C_{3}^{\mathrm{AB}}\, f_{i+3}
  + C_{2}^{\mathrm{AB}}\, f_{i+2}
  + C_{1}^{\mathrm{AB}}\, f_{i+1}
  + C_{0}^{\mathrm{AB}}\, f_{i},
}
where the coefficients are defined as,
\eqarray{
  C_{3}^{\mathrm{AB}}
  &=& \frac{1}{12 (x_i - x_{i+3}) (x_{i+2} - x_{i+3}) (x_{i+1} - x_{i+3})}\nonumber\\
  & & \big[3 x_{i+3}^4 - 4 (x_i + x_{i+1} + x_{i+2}) x_{i+3}^3\nonumber\\
  &+& \{(6 x_i + 6 x_{i+2} ) x_{i+1} + 6 x_i  x_{i+2}\} x_{i+3}^2
  - 12  x_i  x_{i+1}  x_{i+2}  x_{i+3}\nonumber\\
  &+& \{(12 x_i x_{i+4}  -  6 x_{i+4}^2)  x_{i+2}
  -  6 x_i  x_{i+4}^2  +  4 x_{i+4}^3\} x_{i+1}\nonumber\\
  &+& (- 6 x_i  x_{i+4}^2  +  4 x_{i+4}^3)  x_{i+2}
  +  4 x_i  x_{i+4}^3  -  3 x_{i+4}^4\big],\\
  C_{2}^{\mathrm{AB}}
  &=& (x_{i+3} - x_{i+4})^2\nonumber\\
  & & \frac{6  x_i  x_{i+1} - 2  x_i  x_{i+3} - 4  x_i  x_{i+4} - 2  x_{i+1}  x_{i+3}
  - 4  x_{i+1}  x_{i+4} + x_{i+3}^2 + 2  x_{i+3}  x_{i+4} + 3 x_{i+4}^2}
  {12  (x_i - x_{i+2})  (x_{i+1} - x_{i+2})  (x_{i+2} - x_{i+3})},\\  
  C_{1}^{\mathrm{AB}}
  &=& - (x_{i+3}  -  x_{i+4})^2 \nonumber\\
  & & \frac{6 x_i  x_{i+2} - 2  x_i  x_{i+3} - 4  x_i  x_{i+4} - 2  x_{i+2}  x_{i+3}
  - 4  x_{i+2}  x_{i+4}  +  x_{i+3}^2 + 2  x_{i+3}  x_{i+4}  +  3 x_{i+4}^2}
  {12 (x_i  -  x_{i+1})  (x_{i+1}  -  x_{i+3})  (x_{i+1}  -  x_{i+2})},\\
  C_{0}^{\mathrm{AB}}
  &=& (x_{i+3}  -  x_{i+4})^2\nonumber\\
  & & \frac{6 x_{i+1}  x_{i+2} - 2  x_{i+1}  x_{i+3} - 4  x_{i+1}  x_{i+4}
    - 2  x_{i+2}  x_{i+3}
    - 4  x_{i+2}  x_{i+4}  +  x_{i+3}^2  +  2  x_{i+3}  x_{i+4}  +  3  x_{i+4}^2}
       {12  (x_i  -  x_{i+3})  (x_i  -  x_{i+2}) (x_i  -  x_{i+1})}.\nonumber\\
}

We then use the solution obtained from the Adam-Bashforth method
into Adam-Moulton corrector method which can be written as,
\eqarray{
  \label{eq:adam_moulton}
  y_{i+4} &=& y_{i+3} + C_{4}^{\mathrm{AM}}\, f_{i+4} + C_{3}^{\mathrm{AM}}\, f_{i+3} + C_{2}^{\mathrm{AM}}\, f_{i+2} + C_{1}^{\mathrm{AM}}\, f_{i+1},
}
where,
\eqarray{
  C_{4}^{\mathrm{AM}}
  &=& (x_{i+4} - x_{i+3})\nonumber\\
  & & \frac{(6x_{i+1}x_{i+2} - 2x_{i+1}x_{i+3} - 4x_{i+1}x_{i+4} - 2x_{i+2}x_{i+3} - 4x_{i+2}x_{i+4} + x_{i+3}^2 + 2x_{i+3}x_{i+4} + 3x_{i+4}^2)}{12(x_{i+2} - x_{i+4})(x_{i+1} - x_{i+4})},\nonumber\\
  &&\\
  C_{3}^{\mathrm{AM}}
  &=& (x_{i+4} - x_{i+3})\nonumber\\
  & &\frac{(6x_{i+1}x_{i+2} - 4x_{i+1}x_{i+3} - 2x_{i+1}x_{i+4}
    - 4x_{i+2}x_{i+3} - 2x_{i+2}x_{i+4} + 3x_{i+3}^2 + 2x_{i+3}x_{i+4} + x_{i+4}^2)}
  {12(x_{i+2} - x_{i+3})(x_{i+1} - x_{i+3})},\nonumber\\
  &&\\
  C_{2}^{\mathrm{AM}}
  &=&
  \frac{(2x_{i+1} - x_{i+3} - x_{i+4})(x_{i+3} - x_{i+4})^3}
       {12(x_{i+2} - x_{i+4})(x_{i+2} - x_{i+3})(x_{i+1} - x_{i+2})},\\
  C_{1}^{\mathrm{AM}}
  &=& -\frac{(2x_{i+2} - x_{i+3} - x_{i+4})(x_{i+3} - x_{i+4})^3}
  {12(x_{i+1} - x_{i+4})(x_{i+1} - x_{i+3})(x_{i+1} - x_{i+2})}.
}

We then use the Eqs.~\eqref{eq:adam_bashforth}
and~\eqref{eq:adam_moulton} to solve Eq.~\eqref{eq:Schreq}
as described in Sec.~\ref{sec:NUSPINEVOL}.

\vspace*{15mm}
  

\end{document}